\documentclass[11pt,document,nofootinbib,superscriptaddress,onecolumn,preprintnumbers,balancelastpage]{article}
\usepackage{jheppub} 
\usepackage{graphicx}
\usepackage{epstopdf}
\usepackage{dcolumn}
\usepackage{bm}
\usepackage{hyperref}
\usepackage{booktabs}
\usepackage[dvipsnames]{xcolor}
\usepackage{amsmath}
\usepackage{amssymb}
\usepackage{cancel}
\usepackage{xpatch}
\usepackage{fontawesome} 
\usepackage{arydshln}
\usepackage{xspace}
\usepackage{mathtools}
\usepackage{bbold}
\usepackage{booktabs, multirow} 
\usepackage{soul}
\usepackage{tikz}
\usepackage{graphicx} 
\usepackage{physics}
\usepackage{comment}
\usepackage[shortlabels]{enumitem}

\definecolor{deepgreen}{rgb}{0.2,0.8,0.2}

\definecolor{deepblue}{rgb}{0.2,0.4,0.8}

\definecolor{deepred}{rgb}{0.8,0.2,0.2}

\newcommand{\gev}{\text{GeV}\,}
\newcommand{\sm}{\text{SM}\,}

\newcommand{\model}{\text{\rm I(1+2)HDM}\,}

\usepackage{todonotes}

\allowdisplaybreaks

\title{Probing Standard Model-like di-Higgs Production at Photon-Photon Colliders in the I(1+2)HDM Type-I}

\author[a,b]{Abdesslam Arhrib,}
\author[c]{Ayoub Hmissou,}
\author[d,e]{Stefano Moretti,}
\author[c]{and Larbi Rahili}

\affiliation[a]{Facult\'e des Sciences et Techniques, Universit\'e Abdelmalek Essaadi, B. 416, Tangier, Morocco}
\affiliation[b]{LAPTh, CNRS, Universit\'e Savoie Mont-Blanc, 9 Chemin de Bellevue, 74940, Annecy, France}
\affiliation[c]{Laboratory of Theoretical Physics and High Energies (LPTHE), Physics Department, FSA, Ibnou Zohr University, P.O.B. 8106 Agadir, Morocco}
\affiliation[d]{School of Physics and Astronomy, University of Southampton, Southampton, SO17 1BJ, United Kingdom}
\affiliation[e]{Department of Physics and Astronomy, Uppsala University, Box 516, 751 20 Uppsala, Sweden}

\emailAdd{aarhrib@uae.ac.ma}
\emailAdd{ayoub1hmissou@gmail.com}
\emailAdd{stefano.moretti@cern.ch}
\emailAdd{rahililarbi@gmail.com}

\abstract{In this paper, pair production of Standard Model (SM)-like Higgs bosons, $hh$, is studied through $\gamma\gamma$ scattering at future electron-positron colliders, in the framework of the Inert Doublet Model with two Active Doublets, i.e., the I(1+2)HDM for short. The relevance of the process $\gamma\gamma\to hh$ for such a Beyond the SM (BSM) scenario stems from the fact that it is a one-loop process at lowest order, wherein inert charged states $\chi^\pm$ contribute alongside with $W^\pm$, $H^\pm$ and heavy fermions (primarily, bottom and top quarks), crucially, at the same perturbative order. {{Given that $\chi^\pm/H^\pm$ masses and $hS^+S^-$ ($S^\pm=\chi^\pm, H^\pm$) couplings are very mildly constrained,}} there exist regions of the parameter space of the I(1+2)HDM where the former can be rather light and the latter rather large. After imposing up-to-date  theoretical and experimental constraints on the I(1+2)HDM, it is found that the production rates of such process at future $\gamma\gamma$ machines  can be enhanced up to a factor of $\approx$ $50$  with respect to the SM, significantly exceeding typical yields of  conventional 2-Higgs Doublet Models (2HDMs). Further, thanks to the level of control that one can attain at such facilities on the photon  kinematics, leading to excellent invariant mass resolution of the incoming photon pairs, we show how it is possible to extract from this process the value of the $\chi^\pm$ mass (along that of the active $H^\pm$ states) with high precision, whichever the decays of the $hh$ pair, both with and without beam polarization.}

\begin{document}
\maketitle
\flushbottom

\section{Introduction}
\label{sec:introduction}
The Standard Model (SM) particle spectrum has been completed with the discovery of the Higgs boson ($h$) on $4^{th}$ July 2012, by the ATLAS and CMS experiments at the Large Hadron Collider (LHC)  \cite{ATLAS:2012yve,CMS:2012qbp}. Furthermore, this discovery has confirmed the  Higgs mechanism of Electro-Weak Symmetry Breaking (EWSB) and consequent mass generation. The two collaborations also  carried out several Higgs couplings measurements at the LHC Run-I \cite{ATLAS:2016neq} and Run-II \cite{CMS:2018uag,ATLAS:2019nkf}, such as those to fermions and gauge bosons, with uncertainty at most of $30-50$\% and 20\%, respectively. The aforementioned measurements will be improved at future experiments such as LHC Run-III and  the High-Luminosity LHC option (HL-LHC) \cite{Gianotti:2002xx,Apollinari:2015wtw}. Herein, for example, the  $h b \bar b$, $h  \tau^+ \tau^-$ and $h  ZZ$ couplings will be measured with $4$-$7\%$, $2$-$5\%$ and $2$-$4\%$ precision, respectively. In addition, the above experimental uncertainties will be further reduced in the clean environment offered by future $e^+e^-$ colliders, such as the  International Linear Collider (ILC) \cite{Bambade:2019fyw}, the  Circular  Electron-Positron  Collider (CEPC) \cite{CEPC-SPPCStudyGroup:2015csa}, the Future Circular Collider operating in $e^+e^-$ mode  (FCC-ee)  \cite{TLEPDesignStudyWorkingGroup:2013myl,FCC:2025lpp} or else the Compact Linear Collider (CLIC) \cite{CLICPhysicsWorkingGroup:2004qvu,Linssen:2012hp,Adli:2025swq}. Again, for example, at the ILC, the $h b \bar b$, $h  \tau^+ \tau^-$ and  $h  Z Z$ couplings will be measured with $1.5\%$, $1.9\%$ and $0.6\%$ accuracy, respectively \cite{Fujii:2015jha}.

A reason for measuring the couplings of the discovered Higgs state (with mass 125 GeV) is to understand whether these signal the presence of some New Physics (NP) Beyond the SM (BSM), as the SM is plagued with several flaws (see, e.g., Ref.~\cite{Khalil:2022toi} for  a review of these). From the experimental side, the SM is unable to account for neutrino masses, the matter-antimatter asymmetry in the Universe or to provide a candidate for Dark Matter (DM). From the theoretical side, the SM suffers from the hierarchy problem (i.e., the inability to reconcile the EW scale with 
that of gravity without unnatural fine-tuning of its parameters). However, despite a flurry of NP models present in literature, there is  
 no conclusive evidence in data for any of these yet. The end of Run-III at the LHC and the next phase at the CERN machine (the aforementioned HL-LHC) will hopefully bring some signals of BSM physics (or, at least, hints of it), but this is not a certainty. That is, one may need to  wait for the next generation of accelerators. Among the latter, the $e^+e^-$ ones mentioned above have the advantage of providing a very clean environment wherein to search even for subtle hints of NP, compared to hadronic machines, wherein the QCD backgrounds (including those associated to the remnants from the initial state) are formidable.  
Such electron-positron machines also have the option of running in $\gamma\gamma$ mode, thereby effectively being $\gamma\gamma$ colliders, wherein the photon beams are generated from Compton back-scattering of laser light \cite{ECFADESYPhotonColliderWorkingGroup:2001ikq,Telnov:2006cj,DeRoeck:2003cjp,Ginzburg:2020koq,Muhlleitner:2005pr}. Specifically, high-energy electrons/positrons from the main accelerator collide with intense laser beams, effectively converting the electrons/positrons into (similar) high-energy photons. These photons are then directed to interact with each other in a separate (from the one of the $e^+e^-$ beams) Interaction Point (IP) to produce $\gamma\gamma$ collisions.

Of particular interest for our purposes, which include to study the effects of beam polarization, are linear colliders, so that we concentrate here on the ILC and CLIC prototypes, thus with the energy of the photon-photon scattering ranging from 250 to 1400 GeV. This will be instrumental to test the possibility of NP entering the Higgs sector, which can be put under 
intense scrutiny in $\gamma\gamma$ collisions at CLIC, through both precise measurements of the discovered SM-like Higgs state  and  direct searches for BSM Higgs particles \cite{Asner:2001vh}. Aside from the obvious task of providing a direct measurement for the  coupling of the SM-like Higgs  boson to two photons, $\gamma \gamma h$, i.e., at production rather than at decay level,
a
 cornerstone of the physics programme at the $\gamma\gamma$ option of CLIC is the 
exploration of trilinear (or triple) Higgs couplings, through the $\gamma\gamma\to hh$ process. Such a loop-induced channel has been studied within the SM in Refs.~\cite{Jikia:1992zw,Jikia:1992mt,Borden:1993cw,Belusevic:2004pz,Berger:2025ijd} and 
found to be rather small while in scenarios with additional particle content it can be significantly enhanced: e.g., in   
Supersymmetric constructs  like the Minimal Supersymmetric SM (MSSM) and Next-to-MSSM (NMSSM)  \cite{Muhlleitner:2001kw,Asner:2001ia,Heng:2013wia,Zhu:1997nz,Zhou:2003ss} or non-Supersymmetric ones like 2-Higgs Doublet Models  
(2HDMs)~\cite{Cornet:2008nq,Asakawa:2008se,Asakawa:2010xj,Arhrib:2009gg,Hernandez-Sanchez:2011idv,Demirci:2019kop,Phan:2024vfy},
wherein some significant enhancements have been seen in the total cross section because of charged particles from NP entering the loops (i.e., charged Higgs states, $H^\pm$, and new scalars or fermions).  

In this paper, we will study how one  such Higgs sector  extensions, namely, the I(1+2)HDM of Ref.~\cite{Grzadkowski:2009bt}, can alter the phenomenology 
of the $\gamma\gamma\to hh$ process. Such a BSM scenario is well motivated as it 
 provides a viable DM candidate as well as  extra CP-violating phases for the explanation of the Universe's matter-antimatter asymmetry. Furthermore, it avoids the so-called little hierarchy problem (i.e., the fine-tuning required in many NP models, where the natural scale for the corresponding new particles would imply a much heavier Higgs boson than what observed). In fact,
such a BSM scenario has also been found to be a viable explanation of two present anomalies from Higgs data, showing excesses at 
95 and 650 GeV \cite{Hmissou:2025uep,Hmissou:2025riw}.  
In such a NP scenario, after EWSB, one is  left  with a spectrum of 8 physical Higgs bosons, 4 of which are charged (so that they enter the 
$\gamma\gamma\to hh$ process at one-loop): 2 active ($H^\pm$) and 2 inert ($\chi^\pm$) ones. In the end, we will show that the contribution of the latter can be significantly larger than that of the former, so that the cross section for this channel can increase by several orders of magnitude above that of the SM (and also exceed the typical  ones predicted in the aforementioned BSM scenarios.)

This paper is organized as follows. In Section~\ref{sec:model_framwork}, a brief review  of  the  I(1+2)HDM is given, crucially, including formulas for trilinear and 
 quartic Higgs self-couplings, also considering both experimental and theoretical constraints on its parameter space.  
 In Section~\ref{sec:sm}, a discussion of the diagrammatic components of the $\gamma\gamma \to hh$ process is presented, where, 
 after describing our computational framework, we illustrate 
our numerical results. The summary is finally given in Section~\ref{sec:summary}.

%
\section{The I(1+2)HDM Framework}
\label{sec:model_framwork}
In what follows, we briefly review the I(1+2)HDM setup, discuss the main theoretical and experimental constraints on its parameter space and present the trilinear and quartic Higgs self-couplings that are directly relevant to di-Higgs production in photon-photon collisions.

\subsection{Model Setup}
\label{subsec:model_setup}
Our theoretical framework is built upon the I(1+2)HDM of Refs. \cite{Grzadkowski:2009bt,Moretti:2015cwa} (see also \cite{Keus:2013hya}). In this model, alongside the 2HDM active weak doublets, represented as $\Phi_i \sim (1, 2, 1/2)$ ($i=1,2$), an additional inert doublet,  denoted as $\eta \sim (1, 2, 1/2)$, is added. This extension of the 2HDM has been extensively studied and is considered as one of the most well-motivated ones. This primarily stems from the DM problem, since the inert doublet naturally provides a stable candidate protected by a discrete $\mathbf{Z}_2$ symmetry, under which the inert doublet transforms as $\eta \rightarrow -\eta$ while all other fields remain invariant under it. In parallel, a softly broken $\mathbf{Z}^\prime_2$ symmetry is introduced to avoid tree level Flavor Changing Neutral Current (FCNC) processes, under which $\Phi_1 \rightarrow +\Phi_1$ and $\Phi_2 \rightarrow -\Phi_2$.

Assuming Charge/Parity (CP) conservation in the Higgs sector, the most general gauge-invariant and renormalizable scalar potential that respects the above discrete symmetry, $\mathbf{Z}_2 \times \mathbf{Z}^\prime_2$, can thus be
divided into three parts, as follows \cite{Merchand:2019bod,Hmissou:2025uep}:
\begin{equation} \label{Eq:fullpot}
V(\Phi_1,\,\Phi_2,\,\eta) = V(\Phi_1\,,\,\Phi_2) + V(\eta) + V(\Phi_1\,,\,\Phi_2\,,\,\eta).
\end{equation}
The first one describes the 2HDM scalar sector and reads
\begin{eqnarray} 
V(\Phi_1,\Phi_2) &=& -\frac12\left\{m_{11}^2\Phi_1^\dagger\Phi_1 
+ m_{22}^2\Phi_2^\dagger\Phi_2 + \left[m_{12}^2 \Phi_1^\dagger \Phi_2 
+  \text{h.c.} \right] \right\} + \frac{\lambda_1}{2}(\Phi_1^\dagger\Phi_1)^2 
+ \frac{\lambda_2}{2}(\Phi_2^\dagger\Phi_2)^2\nonumber \\
&& + \lambda_3(\Phi_1^\dagger\Phi_1)(\Phi_2^\dagger\Phi_2) 
+ \lambda_4(\Phi_1^\dagger\Phi_2)(\Phi_2^\dagger\Phi_1) + \frac12\left[\lambda_5(\Phi_1^\dagger\Phi_2)^2 + \text{h.c} \right].
\label{v12} 
\end{eqnarray}
The second part corresponds to the inert doublet self-interactions,
\begin{equation}
 V( \eta) = m_\eta^2\eta^\dagger \eta + \frac{\lambda_\eta}{2} 
(\eta^\dagger \eta)^2,
\label{v3} 
\end{equation}
and, finally, the third part encodes the interactions terms between $\eta$ and $\Phi_{1,2}$ fields that are defined as 
\begin{align}
 V(\Phi_1,\Phi_2,\eta) 
&=
\lambda_{11\eta\eta} (\Phi_1^\dagger\Phi_1)(\eta^\dagger \eta)
+\lambda_{22\eta\eta} (\Phi_2^\dagger\Phi_2)(\eta^\dagger \eta) +\lambda_{1\eta\eta1}(\Phi_1^\dagger\eta)(\eta^\dagger\Phi_1) \nonumber  \\
& +\lambda_{2\eta\eta2}(\Phi_2^\dagger\eta)(\eta^\dagger\Phi_2) +\frac{1}{2}\left[\lambda_{1\eta1\eta}(\Phi_1^\dagger\eta)^2 +\text{h.c.} \right]  
+\frac{1}{2}\left[\lambda_{2\eta2\eta}(\Phi_2^\dagger\eta)^2 +\text{h.c} \right],
\label{Active-Inert}
\end{align}
wherein,  without loss of generality, all incorporated dimensionless parameters, $\lambda_{1-5}$, $\lambda_{\eta}$, $\lambda_{ii\eta\eta}$, $\lambda_{i \eta \eta i}$ and $\lambda_{i \eta i \eta}$, with $i=1,2$, are assumed to be real. 

The three complex $SU_L(2)$ doublet scalar fields, $\Phi_1$, $\Phi_2$ and  $\eta$, are defined by:
\begin{equation} 
\Phi_i=\left(
\begin{array}{c}\phi_i^+\\ (v_i+\eta_i+i z_i)/\sqrt{2}
\end{array}\right), \quad
i=1,2 \ ,  \quad
\eta = \left(
\begin{array}{c}
 \chi^{+} \\ (\chi + i \chi_a)/\sqrt{2} 
\end{array}
\right),
\end{equation}
where $v_1$ and $v_2$ refer for the Vacuum Expectation Values ({VEV}s) of the Higgs fields $\Phi_1$ and $\Phi_2$, respectively, 
 
To avoid CP-violation in the Higgs sector, all parameters of the scalar potential are assumed to be real. Moreover, and due to the  $\mathbf{Z}_2$ symmetry, the inert scalar doublet $\eta$ does not mix with the two active ones $\Phi_1$ and $\Phi_2$. Consequently, the physical scalar spectrum splits into two sectors.
\begin{itemize}
\item [$\bullet$] Active sector: equivalent to that of a 2HDM, with two CP-even states ($h$ and $H$), one CP-odd ($A$) and a pair of charged Higgs ($H^{\pm}$), with masses $m_h < m_H$, $m_A$ and $m_{H^\pm}$, respectively. At the tree level, the mass relations of this sector remain the same as in the 2HDM. Furthermore, the mixing between weak and mass eigenstates is controlled by the angles $\beta$ (in the charged and CP-odd sectors) and $\alpha$ (in the CP-even sector). The transitions between the non-physical fields and physical scalars is done as follows:
\begin{align}
&\begin{pmatrix}
\phi_1^\pm \\
\phi_2^\pm 
\end{pmatrix}
= \mathcal{R}_\beta
\begin{pmatrix}
G^\pm \\
H^\pm 
\end{pmatrix},~
\begin{pmatrix}
z_1^{} \\
z_2^{}
\end{pmatrix}
= \mathcal{R}_\beta
\begin{pmatrix}
G^0 \\
A 
\end{pmatrix}\,\text{and}\,\,
\begin{pmatrix}
\eta_1^{} \\
\eta_2^{}
\end{pmatrix}
= \mathcal{R}_\alpha
\begin{pmatrix}
H \\
h 
\end{pmatrix},
\end{align}
where $\mathcal{R}_\theta=\{\{\cos\theta \,,\, -\sin\theta\},\{\sin\theta \,,\, \cos\theta\}\}$ is the usual rotation $2\times 2$ matrix. Here, $G^0$ and $G^{\pm}$ are the neutral and charged Goldstone bosons absorbed as the longitudinal components of the $Z$ and $W^{\pm}$, respectively.\\
Starting with ten real parameters in the scalar potential in Eq.(\ref{v12}), and taking into account the two minimization conditions together with the EW relation that fixes the $W^\pm$ mass, $2\,m_{W^\pm}=g\,v$  (with $v=\sqrt{v_1^2+v_2^2}=246$ GeV), we are left with seven free parameters parameters:
\begin{equation}
\Omega_1 = \left\{ m_h,\, m_A,\, m_H,\, m_{H^{\pm}},\, m_{12}^2,\, \tan\beta,\, \cos(\beta-\alpha) \right\}.
\label{set1}
\end{equation}
\item [$\bullet$] Inert sector: from which four additional scalars arise, namely, $\chi$, $\chi_a$ and $\chi^{\pm}$. For simplicity, we assume that the inert doublet $\eta$ couples symmetrically to both active doublets $\Phi_1$ and $\Phi_2$. This assumption preserves the stability of the inert sector while significantly reducing the number of free parameters in the model. As a result, the analysis becomes more tractable while still preserving the essential features of the I(1+2)HDM. To ensure this, we have implemented the following identifications:
\begin{equation}
\label{Eq:DarkDemocracy}
\lambda_a\equiv \lambda_{11\eta\eta}=\lambda_{22\eta\eta}, \quad \lambda_b\equiv \lambda_{1\eta\eta1}=\lambda_{2\eta\eta2} \quad \text{and}\quad \lambda_c\equiv \lambda_{1\eta1\eta}=\lambda_{2\eta2\eta},
\end{equation}
which leave us with only five extra parameters. Consequently, the squared masses of the inert scalars can be expressed as follows:
\begin{eqnarray}
&& m^2_{\chi^{\pm}} = m_{\eta}^2 + \frac{1}{2} \lambda_{a} v^2, \label{masse2chipm}\\
&& m^2_{\chi} = m_{\eta}^2 + \frac{1}{2} (\lambda_{a} + \lambda_{b} + \lambda_{c})v^2 = m^2_{\chi^{\pm}} + \frac{1}{2} (\lambda_{b} + \lambda_{c})v^2 , \label{masse2chieven}\\
&& m^2_{\chi_a} = m_{\eta}^2 + \frac{1}{2} (\lambda_{a} + \lambda_{b} - \lambda_{c})v^2 = m^2_{\chi^{\pm}} + \frac{1}{2} (\lambda_{b} - \lambda_{c})v^2. \label{masse2chiodd}
\end{eqnarray}
The above couplings $\lambda_{a}$, $\lambda_{b}$ and $\lambda_{c}$ can be expressed using the square masses as follows: 
\begin{eqnarray}
&& \lambda_a = \frac{2 (m_{\chi^\pm}^2 - m_{\eta}^2)}{v^2}, \\
&& \lambda_b = \frac{m_{\chi_a}^2 - 2 m_{\chi^{\pm}}^2 + m_{\chi}^2}{v^2}, \\
&& \lambda_c = \frac{-m_{\chi_a}^2 + m_{\chi}^2}{v^2}.
\end{eqnarray}
We finally take the inert physical parameter basis defined in terms of the following 5 inputs:
\begin{equation}
\Omega_2 = \left\{  m_{\chi},\, m_{\chi_a},\, m_{\chi^{\pm}},\, m_{\eta}^2,\, \lambda_{\eta} \right\}.\label{set2}
\end{equation}
\end{itemize}

Thus, altogether, the \model parameter space will be described by the 12 independent  parameters given by $\Omega = \Omega_1 + \Omega_2$.

Before concluding this section, let us note that the presence of the inert doublet does not affect the interactions with either fermions or gauge bosons. Indeed, since $\eta$ is odd under the imposed $\mathbf{Z}_2$ symmetry, it does not couple directly to fermions and/or gauge fields. As a result, the Yukawa and gauge structures of the I(1+2)HDM remain identical to those of the conventional 2HDM and the couplings of the physical Higgs bosons $h$, $H$, $A$ and $H^\pm$ to both fermions and gauge bosons are unchanged \cite{Branco:2011iw}. However, as we shall see (specifically, here, for the case of the charged states), the   $\chi$, $\chi_a$ and $\chi^\pm$ field can enter at loop level. 
In what follow, for the active Higgs sector we assume that we have 2HDM Yukawa texture of the type I where only the second doublet $\Phi_2$ interacts with all the fermions \cite{Branco:2011iw}.

\subsection{Theoretical and Experimental Constraints}
\label{subsec:theo+exp+constraints}
To assess the phenomenology of the model, the above parameters of the \model scalar potential are scanned randomly over the following ranges:
\begin{eqnarray}
&& m_{h} = 125.09\,\gev,\,\, m_H \in [130,\, 10^3]\, \gev,\,\, m_A,m_{H^\pm} \in [100,\, 10^3]\, \gev,\,\,  m_{\chi},m_{\chi_a},m_{\chi^{\pm}}\in [80,\, 10^3]\,\gev \nonumber\\
&&  m_{12}^2 \in [0,\, 10^6]\, \gev^2,\,\,m_{\eta}^2 \in [-10^6,\, 10^6]\, \gev^2,\,\, \lambda_{\eta} \in [0,\, 4\pi] ,\,\, \tan\beta \in [2,\, 12] \,\,\text{and} \,\, \sin(\beta-\alpha) \in [0.96,\, 1],\nonumber
\end{eqnarray}
and have to satisfy the following constraints.\\[1ex]
\noindent
{Firstly,} we require the followings theoretical constraints to be met.
\begin{itemize}
\item [$\bullet$] Perturbativity: ensures that all quartics couplings $\lambda_i$ in the scalar potential should be restricted to be $\le 4\pi$ to prevent non-perturbative behaviors \cite{Moretti:2015cwa}. 
\item [$\bullet$] Vacuum stability: guarantees the Boundedness From Below (BFB) for the Higgs potential \cite{Grzadkowski:2009bt}. 
\item [$\bullet$] Tree level perturbative unitarity: which must be preserved in all the $2 \rightarrow 2$ scalar scattering processes involving scalars and/or gauge bosons \cite{Moretti:2015cwa}.
\end{itemize}

\noindent
{Secondly,} we scrutinize the remaining sample against the following precision and search bounds.
\begin{itemize}
\item [$\bullet$] Oblique Parameters: we investigate to what extent the $S$, $T$ and $U$  precision observables~\cite{Peskin:1991sw} may constrain  the \model parameter space, and, thus, the scalars involved. In general, these parameters receive contributions from both active and inert doublets \cite{Grimus:2008nb,Arhrib:2012ia,Belyaev:2016lok,Abouabid:2023cdz} and we require $\chi^2_{ST} < 5.99$ for consistency with the current best-fit values \cite{ParticleDataGroup:2020ssz}:  
\begin{eqnarray}
S = 0.05 \pm 0.08, \quad
T = 0.09 \pm 0.07 \quad {\rm and} \quad \lambda_{ST} = 0.92,
\end{eqnarray}
while assuming $U=0$.
\item [$\bullet$] Constraints from LEP-I \cite{ParticleDataGroup:2010dbb}, as follows:
\begin{eqnarray}
&& m_{\chi} + m_{\chi^{\pm}} > m_{W^\pm} , \quad m_{\chi_a} + m_{\chi^{\pm}} > m_{W^\pm},  \\
&& m_{\chi} + m_{\chi_a} > m_{Z} , \quad 2\,m_{\chi^{\pm}} > m_{Z}, \\
&& m_{\chi} < 80\,\gev , \quad m_{\chi_a} < 80\,\gev , \quad m_{\chi_a}-m_{\chi} > 8\,\gev.
\end{eqnarray}
\item [$\bullet$] Constraints from LEP-II \cite{Pierce:2007ut,ALEPH:2013htx,Arbey:2017gmh}, as follows: 
\begin{equation}
\label{eq:lepconstraint2}
m_{\chi^{\pm}} > 70\,\gev.
\end{equation}
\item [$\bullet$] DM search limits from relic density as well as (in)direct detection constraints via {\tt micrOMEGAs} \cite{Alguero:2023zol}.
\item [$\bullet$] Flavor constraints using the public {\tt C++} code {\tt SuperIso} \cite{Mahmoudi:2008tp}, by considering the most sensitive FCNC processes such as: BR$(B \to X_s \gamma)$ \cite{HFLAV:2016hnz}, BR$(B_s \to \mu^+\mu^-)$ \cite{LHCb:2021awg,LHCb:2021vsc} plus BR$(B \to \tau\nu)$\cite{HFLAV:2016hnz}.
\end{itemize}

\noindent
{Lastly,} in order to test compatibility of the active Higgs sector with collider data, we proceeded as follows. 
\begin{itemize}
\item
We used  \texttt{HiggsTools}  \cite{Bahl:2022igd}, embedding \texttt{HiggsSignals-3} \cite{Bechtle:2020uwn} and \texttt{HiggsBounds-6} \cite{Bechtle:2020pkv,CMS:2024ulc}, to test the I(1+2)HDM parameter space against both precision measurements of the 125 GeV Higgs boson and exclusion limits from additional Higgs searches at LEP, Tevatron and the LHC. We finally retained only the parameter points satisfying $\chi^2_{125}-{\rm min}\big(\chi^2_{125}\big) \le {\rm 6.18}$, corresponding to a 95\% Confidence Level (CL).
\end{itemize}

\subsection{Triple and Quartic Higgs Self-couplings Relevant for $hh$ Production}
As mentioned in Section~\ref{sec:introduction}, the study of the di-Higgs production processes at high-energy photon-photon colliders provides a unique opportunity to scrutinize BSM scenarios and  probe the structure of its scalar potential. In particular, such processes are highly sensitive to the trilinear and quartic Higgs self-interactions, which encode direct information about the scalar potential responsible for EWSB.

In the SM, these couplings are completely fixed in terms of the Higgs boson mass $m_h$ and the VEV $v$, leading to the well-known expressions listed in Tab.~\ref{trilinear_quartic_couplings}. (Notice, however, that the quartic coupling of the SM, $hhhh$, does not enter our process.)  It is also well known that $\gamma \gamma \to hh$ is the SM is rather small and difficult to be measured directly, hence, 
we explore here  how the additional scalar states within the I(1+2)HDM can significantly alter the structure and magnitude of such SM interactions and produce  others.

The corresponding scalar spectrum of the I(1+2)HDM introduces new trilinear and quartic couplings among the Higgs fields,  many of which appear at the same perturbative order as the SM ones in loop-induced processes such as $\gamma\gamma \to hh$. Hence, taking into account the theoretical considerations above, the modified and additional trilinear and quartic Higgs self-interactions added to the SM can be extracted from the scalar potential as follows: 
\begin{equation}
\mathcal{V}=\lambda_{ijk}\,h_i h_j h_k+\lambda_{ijkl}\,h_i h_j h_k h_l+\cdots,
\end{equation}
where $\lambda_{ijk}$ and $\lambda_{ijkl}$ denote the effective trilinear and quartic couplings in the mass basis, which are function of the underlying Lagrangian parameters $\lambda_{1-5}$, $\lambda_{a}$ as well as the mixing angles $\alpha$ and $\beta$. At tree-level, the corresponding couplings can be written down as shown in Tab.~\ref{trilinear_quartic_couplings}, where the shorthand notations $c_{\theta}$ and $s_{\theta}$ stand, respectively, for $\cos(\theta)$ and $\sin(\theta)$, while $\lambda_{345}=\lambda_3 + \lambda_4 + \lambda_5$.

%
\begin{table}[!ht]
\centering
\caption{Trilinear $h_ih_jh_k$ and quartic $h_ih_jh_kh_l$ couplings in the \sm and \model (wherein the $h$ is the discovered SM-like Higgs state) entering the $\gamma\gamma\to hh$ process.}
\renewcommand{\arraystretch}{1.25}
\resizebox{\textwidth}{!}{%
\begin{tabular}{l|l|l}
\hline
\hline
\textbf{Model}  &  \textbf{Couplings}    &   \textbf{Expressions}   \\
\hline
SM  & $hhh$ & $-3 m_H^2 / v$ \\
\cite{Djouadi:2005gi}   & $hhhh$ & $-3 m_H^2 / v^2$ \\
\hline
\model  & $hhh$ & $-3\big[ \big( 2 \, c_{\alpha+\beta} + s_{2\alpha} s_{\beta-\alpha} \big) m_h^2 - 2 \, c_{\alpha+\beta} c_{\beta-\alpha}^2 m_{12}^2 \big]\big/s_{2\beta}\big/v $ \\
& $Hhh$ & $-\big[ \big( 2 \, m_h^2 + m_H^2 -3 \, m_{12}^2 \big) \, s_{2\alpha} + m_{12}^2 \, s_{2\beta} \big]\,c_{\beta-\alpha}\big/s_{2\beta}\big/v $ \\
& $h H^\pm H^\mp$ & $-\big[ \big( 3 \, m_h^2 + 2\,m_{H^\pm}^2 -4 \, m_{12}^2 \big) \, c_{\alpha+\beta} + \big(m_h^2 - 2\,m_{H^\pm}^2\big) \big( c_{\alpha+\beta}-2\,s_{2\beta}\,s_{\beta-\alpha}  \big) \big]\big/s_{2\beta}\big/(2v) $ \\
& $h \chi^{\pm}\chi^{\mp}$ & $-2\big[m_{\chi^\pm}^2-m_{\eta}^2\big]\, s_{\beta - \alpha}\big/v$ \\
& $H H^\pm H^\mp$ & $-\big[ \big( 3 \, m_H^2 + 2\,m_{H^\pm}^2 -4 \, m_{12}^2 \big) \, s_{\alpha+\beta} + \big(m_H^2 - 2\,m_{H^\pm}^2\big) \big( s_{\alpha+\beta}-2\,s_{2\beta}\,c_{\beta-\alpha}  \big) \big]\big/s_{2\beta}\big/(2v) $ \\
& $H \chi^{\pm}\chi^{\mp}$ & $-2\big[m_{\chi^\pm}^2-m_{\eta}^2\big]\, c_{\beta - \alpha}\big/v$ \\
\hline
& $h h H^{\pm} H^{\mp}$ & $-\big[m_h^2 c_\alpha^{4} t_\beta^{-2} + 2\, m_{H^\pm}^2 c_\beta^{2} s_\alpha^{2} + c_\alpha s_\alpha \big(4\,m_{12}^2 - (m_h^2 - m_H^2) t_\beta^{-1} s_\alpha^{2} \big)$ \\
&  & $- m_{H^\pm}^2 s_{2\alpha} s_{2\beta}
 - \big(m_h^2 - m_H^2\big) c_\alpha^{3} s_\alpha t_\beta
 - s_\alpha^{2} \big(m_{12}^2 + m_{12}^2 t_\beta^{-4} - m_h^2 t_\beta^{-1} s_\alpha^{2} \big) t_\beta^{3}$ \\
&  & $+ c_\alpha^{2} \big(2 m_{H^\pm}^2 s_\beta^{2}
   - (1 + t_\beta^{-4})(m_{12}^2 t_\beta - m_H^2 s_\alpha^{2}) t_\beta^{2}
   \big)
\big]\big/v^{2}$ \\
&  & $+m_{12}^2 s_{2\,\alpha}\,s_{2\,\beta} - m_H^2 c_\alpha^{4} t_\beta^{2}  + s_\alpha \, c_\alpha \big( (m_h^2-m_H^2) s_\alpha^{2} t_\beta -  4\, m_{H^\pm}^2 s_\beta \, c_\beta \big) $ \\
&  & $+ c_\alpha^{2} \big( c_\beta^{2} (m_{12}^2 - 2 m_{H^\pm}^2) + t_\beta^2 ( m_{12}^2 s_\beta^2 - s_\alpha^2 (1+t_\beta^{-4}) m_h^2 ) \big)  \Big]\Big/v^{2}$ \\
& $h h \chi^{\pm}\chi^{\mp}$ & $-2\big[m_{\chi^\pm}^2-m_{\eta}^2\big]\big/v^{2} $ \\
\hline
\hline
\end{tabular}%
}
\label{trilinear_quartic_couplings}
\end{table}

Importantly, these couplings exhibit two key features. Firstly, the triple, $hhh$, and quartic, {{ $ hhhh$\footnote{Notice that this coupling 
does not enter our process of interest.}}}, Higgs self-couplings deviate from their SM values due to the mixing of the light $h$ state (the SM-like one) with the heavy CP-even Higgs $H$. Secondly, the emergence of new vertices such as $h H^\pm H^\mp$, $h \chi^\pm \chi^\mp$, $h h H^\pm H^\mp$ and $h h \chi^\pm \chi^\mp$, which have no analogues in the SM, provide distinctive signatures that can strongly enhance or suppress the di-Higgs production rates, depending on the \model parameter choices. In particular, the couplings involving inert charged scalars $\chi^\mp$ are of great phenomenological interest, since their size is governed by the parameter $\lambda_{a}$, which remains relatively unconstrained by current data, making it a smoking-gun feature of the I(1+2)HDM framework.

\section{Di-Higgs Production}
\label{sec:sm}
Before analyzing the predictions of the I(1+2)HDM, it is instructive to briefly recall the structure of the process $\gamma\gamma \to hh$ in the SM. This will serve as a reference for identifying possible deviations induced by the extended scalar sector. In particular, we outline the loop-induced nature of the process, the relevant virtual particles involved as well as the computational framework used to obtain the amplitudes. In this context, it is worth noticing that our numerical analyses were performed using the following set of parameters: 
$m_t=171.4$ GeV, $m_b=4.75$ GeV, $m_Z=91.187$ GeV and  $m_{W^\pm}=80.39$ GeV. 
The Weinberg angle $s_W\equiv \sin\theta_W$ is defined in the on-shell scheme as $s_W^2=1-m_W^2/m_Z^2$ and  for the fine structure constant we use $\alpha=1/137.035989$ in the Thomson limit. Subsequently, the one-loop amplitudes are generated by the public code FeynArts \cite{Hahn:2000kx},   written in the Feynman gauge and computed at Leading Order (LO) using FormCalc  \cite{Hahn:2001rv} adopting  dimensional regularization. The outputs are generated in terms of Passarino-Veltman scalar integrals \cite{Passarino:1978jh} which are computed using the LoopTools package \cite{Hahn:1998yk}. We have checked that the total amplitude is Ultra-Violet (UV) finite and also renormalization scale independent, all this providing a good check of our  calculations.
{{The cross section  for $e^+ e^- \to \gamma\gamma \to hh$  is obtained by convolution of the Compton backscattered photon spectra with the partonic 
 $\gamma\gamma \to hh$ cross section. The photon spectra are  taken from the CompAZ library \cite{ECFADESYPhotonColliderWorkingGroup:2001ikq,Ginzburg:1999wz,Ginzburg:1981vm,Ginzburg:1982yr,Telnov:1989sd}, which provides the photon energy spectrum for diverse beam energies, as well as the average photon polarization for any photon energy.}}

\begin{figure}[!t]
\centering
\includegraphics[width=0.86\textwidth]{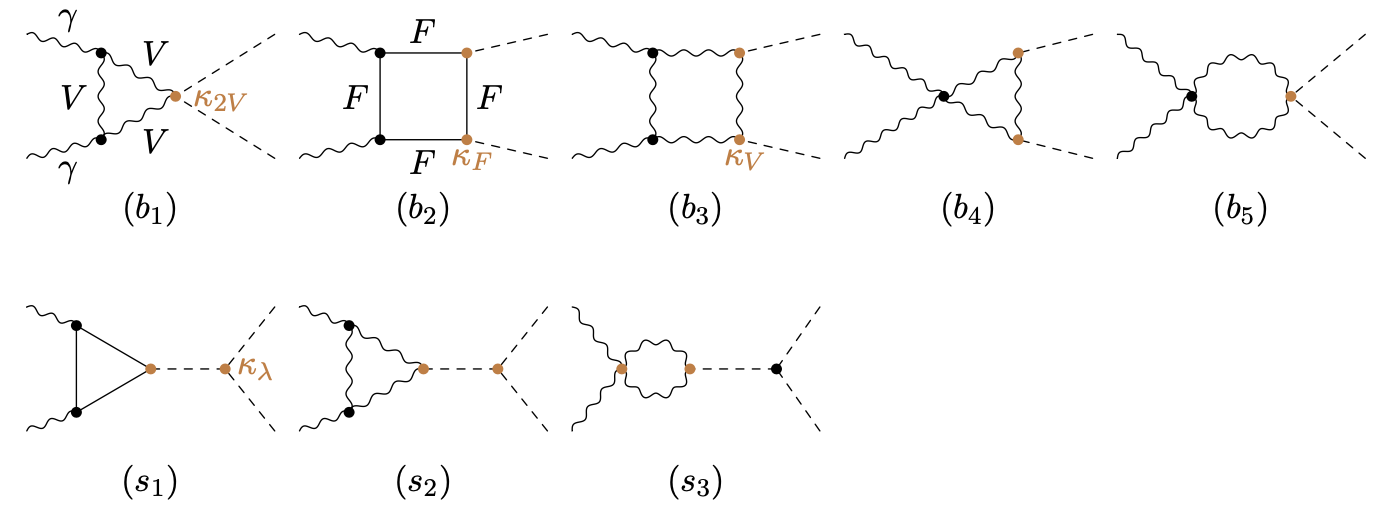}
\caption{Generic Feynman diagrams for di-Higgs production, in the unitary gauge, at a photon-photon collider within the SM. The incoming wavy lines 
correspond to photons while  outgoing dashed lines denote (SM-like) Higgs bosons. The loops are mediated either by fermions $F$ (top, bottom quarks and  FP ghosts, solid lines) or gauge bosons $V$ ($W^\pm$, wavy internal lines).  
 Herein, we distinguish the diagrams in terms of their propagators, as $s$-channels ones ($s_{1-3}$) and others
($b_{1-5}$), hereafter, denoted by `Others (in plots)'. Black and brown blobs refer here to the reduced couplings $\kappa_{V,F,\lambda}$ or $\kappa_{2V}$.}
\label{fig:generic_FD_SM}
\end{figure}

\subsection{The $\gamma\gamma \to hh$ Process in the SM}
In the SM,  the process $\gamma\gamma \to hh$ is mediated at one-loop and receives contributions from  $s$-channel  diagrams $(s_{1-3})$ and the graphs $(b_{1-5})$  as one can see in Fig.~\ref{fig:generic_FD_SM}. (We stress here that we present the Feynman diagrams in the unitary gauge  in order to avoid showing all topologies with charged Goldstone but the calculation is performed in the Feynman gauge using dimensional regularization.) The virtual particles that contribute to this process, as generically depicted therein,  involve: exchange of fermions (quarks and leptons), $W^\pm$ gauge bosons,  $G^\pm$ charged Goldstones and the charged Fadeev-Popov (FP) ghosts. The contribution of light quarks and charged leptons is, in general, negligible because proportional to fermion masses. It is clear that the process $\gamma\gamma \to hh$ is sensitive  to several reduced couplings such as $\kappa_{V, t, \lambda}$ and $\kappa_{2V}$ which are of phenomenological interest. However, given that $\kappa_{V}$ and $\kappa_{t}$ are now quite well measured at the LHC and consistent with the SM predictions, we will address the sensitivity  to $\kappa_{\lambda}$ in what follows.
to all other remaining diagrams.

\begin{figure}[!h]
\centering
\includegraphics[scale=0.4]{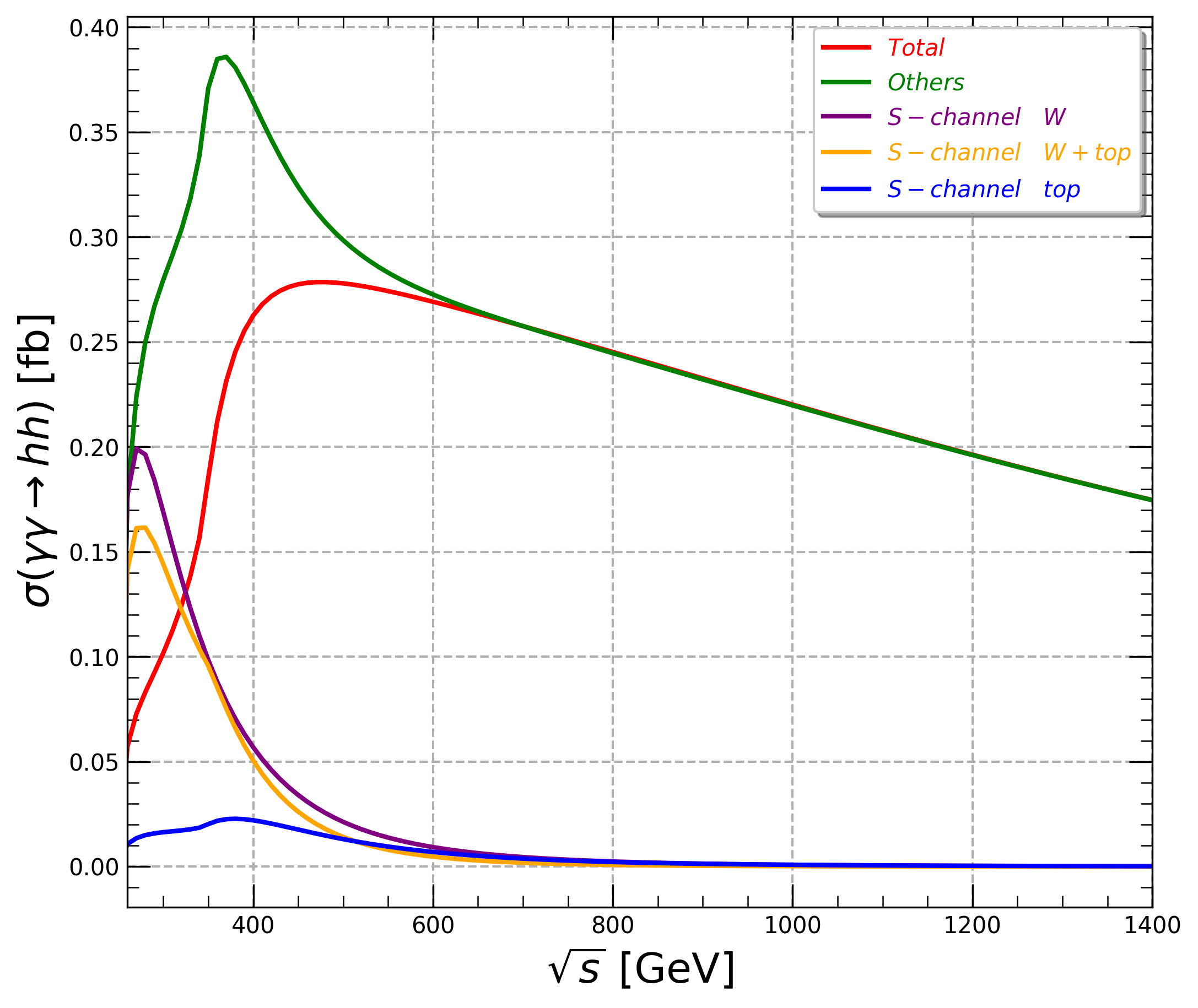}
\includegraphics[scale=0.4]{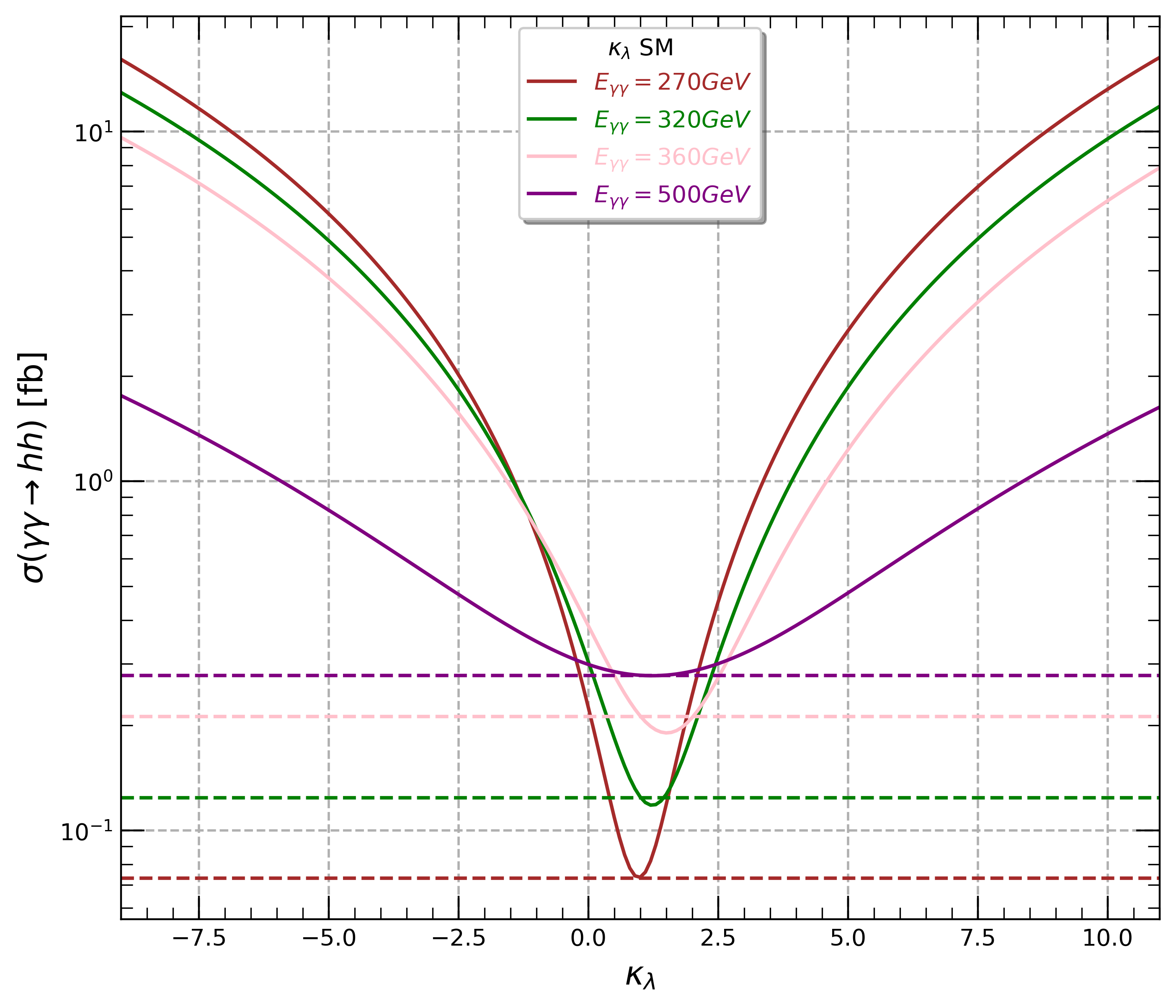}
\caption{Left: cross section for $\gamma\gamma \to hh$  in the SM as a function of the CoM energy. 
The total rate (solid red) is decomposed into two contributions:
$s$-channel loops (solid orange) and other loops (solid green). The contribution from the $s$-channel loops is further decomposed into the ones from top quarks (solid blue) and from $W^\pm$ bosons (solid purple). Right: total cross section for $\gamma\gamma \to hh$ as a function of $\kappa_\lambda$ for several CM energies. The horizontal lines are the SM cross sections ($\kappa_\lambda=1$).}
\label{fig:crosssection_SM}
\end{figure}

We first analyze the total cross section of the partonic process $\gamma\gamma \to hh$ as a function of the $\gamma\gamma$ Center-of-Mass (CoM) energy $\sqrt{s}$, while decomposing the full result in all the involved contributions, from where one can see that the the terms other than the $s$-channel graphs almost entirely dominate  (in turn led by the $W^\pm$ loops over the top ones), as seen in Fig.~\ref{fig:crosssection_SM} (the additional $s$-channel constributions are negligible). 
Hence, at low CoM energy,  a pronounced destructive interference pattern between the SM contributions is maximized, leading to a tremendous peak-dip structure, as can be noted  from the left side of Fig.~\ref{fig:crosssection_SM}. This happens near the $t\bar{t}$ threshold, so that the diagrams contributing significantly towards this trend are those with loops of  top (anti)quarks (i.e., diagrams $(s_{1})$ and $(b_{2})$ with top as internal particle). Typically, the total cross section, exhibited through a red line, increases rapidly after the $t\bar{t}$ threshold and reaches its maximum at a collision energy of $\sqrt{s} = 450-500$~GeV. After that, it  slightly decreases at high energies, where the $b_{1-5}$ topologies dominate the $s_{1-3}$ ones.
In the right panel of Fig.~\ref{fig:crosssection_SM} we illustrate the total cross section of $\gamma\gamma \to hh$  as a function of $\kappa_\lambda$, which is defined as 
$hhh = \kappa_\lambda \times (hhh)^{\rm SM}$, for several CoM energies. By doing so we test  how the NP contribution can modify the partonic cross section $\sigma(\gamma\gamma \to hh)$ through radiative corrections that may affect the trilinear Higgs coupling. 
The SM value is obtained for $\kappa_\lambda=1$ and is represented by a horizontal line for different energies. 
At $\sqrt{s}=320$ and 360 GeV,  in the case of $\kappa_\lambda <1$,  there is an enhancement while, for the case $1<\kappa_\lambda <1.5$, there can
be a  small suppression of the SM cross section.  For energies around 270 GeV and 360 GeV, the cross section is enhanced for all values of $\kappa_\lambda$. 
One can then conclude that the optimal CoM energy to study  di-Higgs production at photon-photon colliders would be 270 GeV or 450 GeV (and above).
Such a $\kappa_\lambda$ is being severely constrained from di-Higgs searches at the LHC via  $gg\to hh$ 
\cite{ATLAS:2024ish,ATLAS:2022jtk,CMS:2024awa} to be in the range  $-1.6 < \kappa_\lambda < 7.2 $ when all couplings are SM-like except for $hhh$. 
It is clear that, by scaling $hhh$ through a $\kappa_\lambda$ factor, the total cross section is enhanced with respect to the SM yield by more than one order of magnitude, 
particularly for a  low CoM energy, between $\sqrt{s}=270$ and 360 GeV.  {{This strong dependence of the cross section $\gamma \gamma \to hh$ on $\kappa_h$
is due to the fact that 
the $s$-channel contribution (from $W^\pm$ and top quark loops) is rather important only before the opening of top threshold at $\sqrt{s}=350$ GeV, so that  
crossing this latter results in a sharp decrease in the $s$-channel contribution and hence a mild enhancement of the cross section as a function of $\kappa_h$.}}


%
%


In the analysis that follows, we investigate how the I(1+2)HDM can alter the $\gamma\gamma\to hh$ process compared to the SM expectations. Herein, the value of $\kappa_\lambda$ is a derived quantity emerging from the $\lambda_i$'s in the corresponding scalar potential after EWSB. However, alongside the 
$\kappa_\lambda$ rescaling effect, 
there will be new Feynman diagrams to consider,   chiefly, those involving new charged scalars (both active and inert) entering the loops, in turn triggering new trilinear, i.e., $hS^+S^-$, and  {{ (now also) quartic, i.e., $hhS+S^-$, self-couplings involving charged scalars, $S^\pm=\chi^\pm,H^\pm$}}. Furthermore, the interactions of the SM-like Higgs boson  will be modified due to the mixing with other (active) scalars by the following scaling factors:
\begin{align}
\kappa_X=\frac{g_{hXX}^{\rm I(1+2)HDM}}{g_{hXX}^{\rm SM}}\,\, \text{with}\,\, X=V \, \text{or}\, F, \quad  \quad 
\kappa_{2V}=\frac{g_{hhVV}^{\rm I(1+2)HDM}}{g_{hhVV}^{\rm SM}}, \quad \quad 
\kappa_{\lambda}=\frac{g_{hhh}^{\rm I(1+2)HDM}}{g_{hhh}^{\rm SM}},
\label{kappa}
\end{align}
denoted as  black and brown blobs   in diagrams $(s_{1-3})$ and $(b_{1-5})$  of  Fig~\ref{fig:generic_FD_SM}.

\subsection{I(1+2)HDM Results}
\label{sec:results}
\begin{figure}[!t]
\centering
\includegraphics[width=0.86\textwidth]{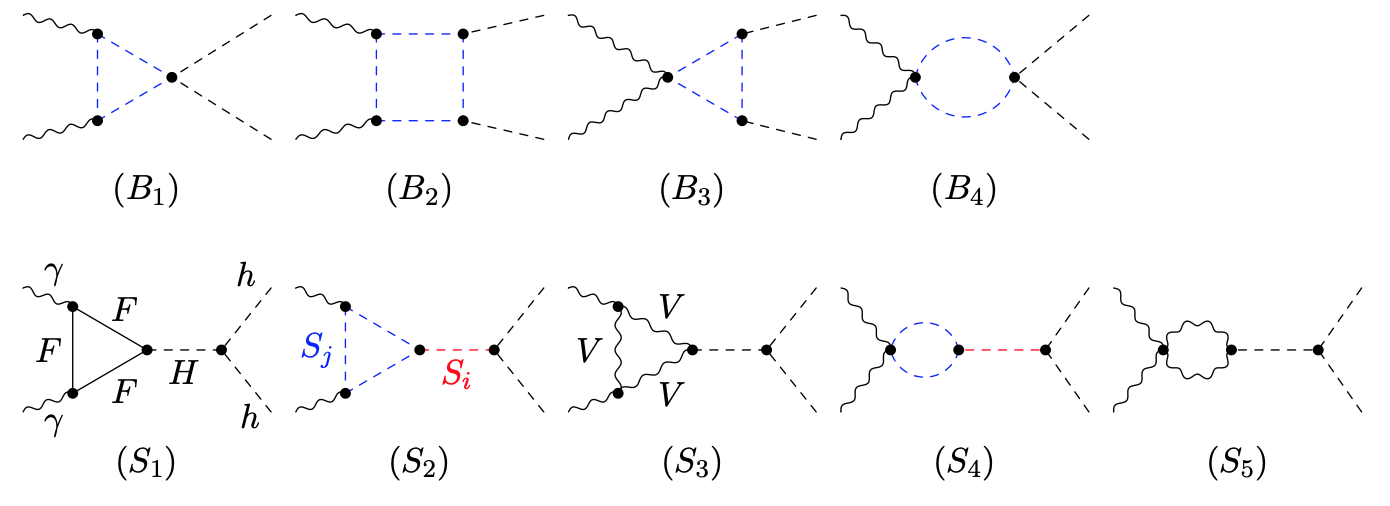 }
\caption{
Generic additional Feynman diagrams for di-Higgs production at a photon-photon collider within the I(1+2)HDM with respect to the SM. Here, we adopt similar graph structures as those introduced for the SM: the $s$-channel ones ($S_{1-5}$), where the internal black(red)[blue] dash lines represent $H$($S_i=h,H$)[$S_j=\chi^\pm,H^\pm$] plus other contributions ($B_{1-4}$).}
\label{fig:generic_FD_I3HDM}
\end{figure}

In this subsection, we analyze the  results for the production process $\gamma\gamma \to hh$ in our NP scenarion, the I(1+2)HDM. To start with, we exhibit in Fig.~\ref{fig:generic_FD_I3HDM} the additional Feynman diagrams that contribute to  Higgs pair production in $\gamma\gamma$ collisions,  originating either from the active  sector, represented by $H$ and $H^\pm$, or from the inert one, denoted by $\chi^\pm$. 


\begin{figure}[!ht]
\centering
\includegraphics[scale=0.4]{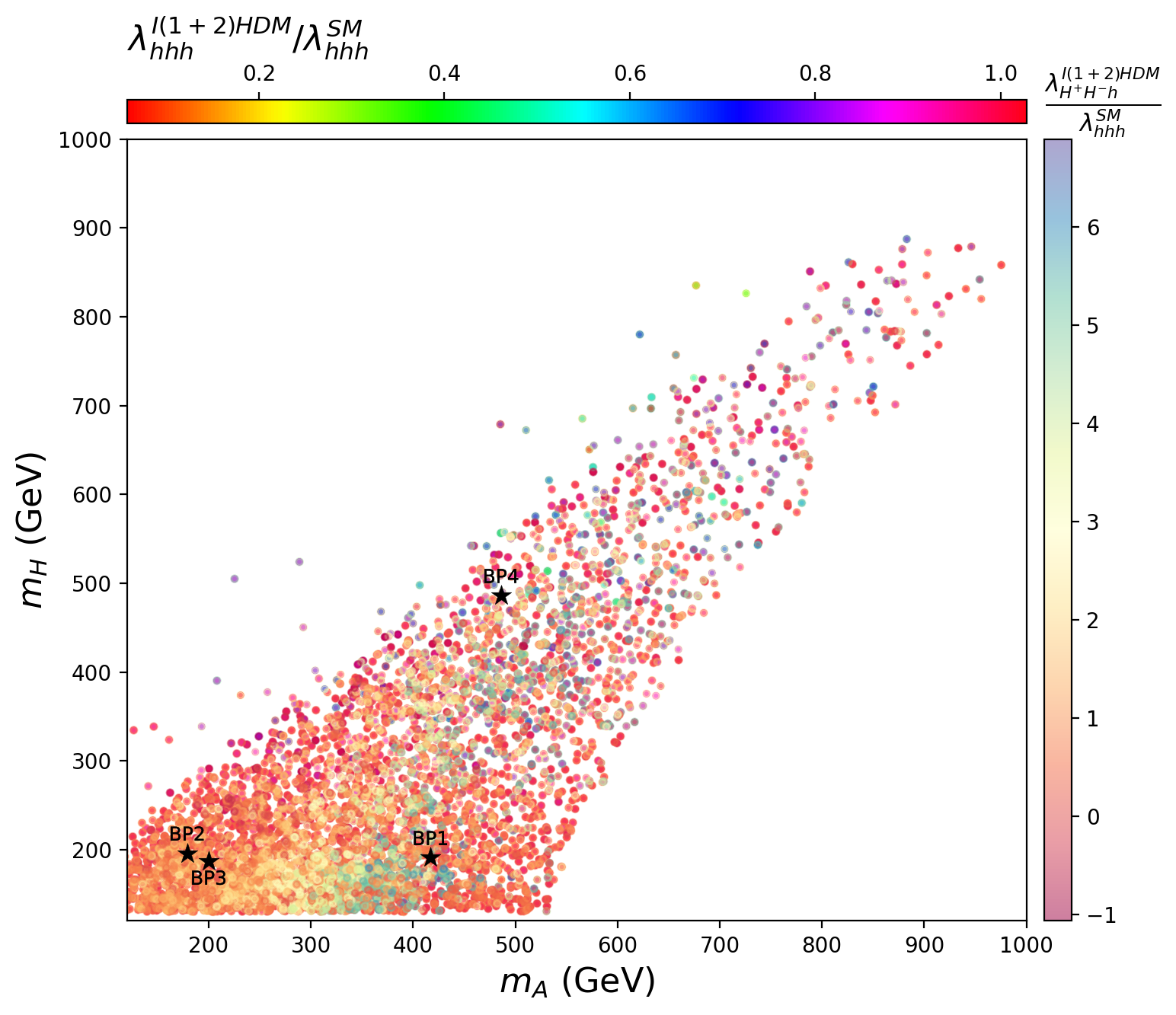}
\includegraphics[scale=0.4]{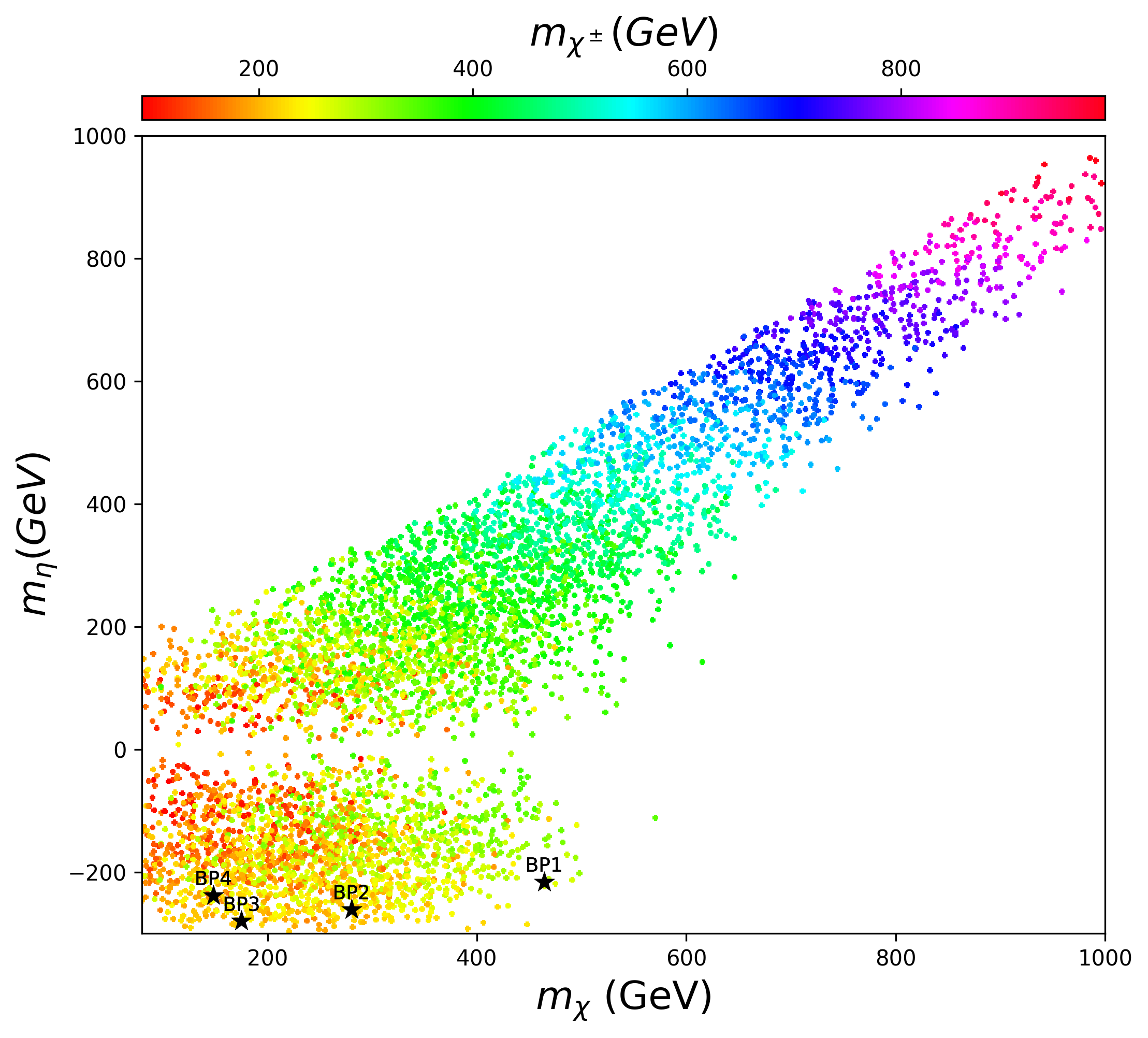}
\caption{Allowed parameter space in the  I(1+2)HDM: left  in the $(m_H,m_{A})$ plane and right in the  $(m_{\chi},m_{\eta})$ plane. We also show 
the reduced couplings $(hhh)^{\rm I(1+2)HDM}/(hhh)^{\rm SM}$ and $(hH^+ H^-)^{\rm I(1+2)HDM}/(hhh)^{\rm SM}$. }
\label{fig:spcparam}
\end{figure}

To illustrate the phenomenological  impact of the new Feynmans diagrams and all rescaled couplings, we conduct an extensive scan of the I(1+2)HDM parameter space according to the intervals mentioned in Subsection~\ref{subsec:theo+exp+constraints}, while we further assume that $\chi$ is the lightest particle, i.e., $m_{\chi} < m_{\chi_{a}}$ and $m_{\chi} < m_{\chi^{\pm}}$, while taking into account all the constraints described above. 
Fig.~\ref{fig:spcparam} illustrates the correlations among $m_A$ and $m_H$ (left panel) and between $m_\chi$ and $m_\eta$ (right panel), with color-coding according to $m_{H^\pm}$ and $m_{\chi^\pm}$, respectively. It is clear that the pattern of correlations is nearly the same between the inert and active sectors plus  both contributions share the feature that a light charged scalar with $m_{H^\pm},\,m_{\chi^\pm} \lesssim 500-600$ GeV generally implies that the neutral states $H$, $A$ and $\chi$ are also quite light, with masses not exceeding  $700-800$ GeV. In the same figure (left), we also present the size of the  $(hH^+ H^-)^{\rm I(1+2)HDM}$ coupling normalized to the SM triple coupling, $(hhh)^{\rm SM}$. One can then see that the triple coupling $(hH^+ H^-)^{\rm I(1+2)HDM}$ could be up to  6 times larger that the corresponding SM value. Furthermore, in some corner of the parameter space, $(hH^+ H^-)^{\rm I(1+2)HDM}$ could be of opposite sign with respect to the SM value. {{We emphasize here the significance of the charged scalars  ($H^\pm$ and $\chi^\pm$)  in the box diagrams $B_2$ and $B_3$. In fact, 
the coupling of the SM-like Higgs state to such particles enters the amplitude of $B_2$ and $B_3$  in quadrature, i.e., 
proportionally to $(hH^+ H^-)^2$ and $(h\chi^+ \chi^-)^2$. 
In contrast, the charged scalar contributions to $S_2$ and $S_4$ are proportional to $(hS^+S^-)\times (hhh)$
or $(HS^+S^-)\times (Hhh)$,  ($S^\pm=\chi^\pm$ or $H^\pm$),  respectively, which are suppressed by a small $\cos(\beta-\alpha)$. }} 

In the  light of these results, we select four Benchmark Points (BPs)  based on the mass hierarchy between the new (both active and inert) scalar  bosons, as follows. 
\begin{itemize}
\item [$\bullet$] {BP1}: $m_H < m_{\chi^\pm} < m_{H^\pm}$,
\item [$\bullet$] {BP2}: $m_H < m_{\chi^\pm} \approx m_{H^\pm}$ (degenerate charged states),
\item [$\bullet$] {BP3}: $m_{H^\pm} < m_H < m_{\chi^\pm}$,
\item [$\bullet$] {BP4}: $m_{\chi^\pm} < m_H < m_{H^\pm}$,
\end{itemize}
which parameters are given in Tab.~\ref{BPs_descriptions}.

\begin{table}[!ht]
\centering
\resizebox{\textwidth}{!}{%
\begin{tabular}{c|c|c|c|c|c|c|c|c|c|c|c|c}
\hline
BPs & $m_h$ & $m_H$ & $m_A$ & $m_{H^\pm}$ & $m_{12}^2$ & $m_{\chi}$ & $m_{\chi_a}$ & $m_{\chi^\pm}$ & $m_{\eta}^2$ & $\lambda_{\eta}$ & $\tan\beta$ & $\sin(\beta-\alpha)$ \\
\hline
{BP1} & \multirow{4}{*}{125.09} & 191 & 417 & 327 &  5625  & 464 & 324 & 290 & $-46947$ & 4.33 & 2.73 & 0.993 \\
{BP2} &  & 196 & 179 & 229 &  4650  & 280 & 155 & 222 & $-68469$ & 4.06 & 3.08 & 0.997 \\
{BP3} & & 187 & 196 & 110 &  8124  & 175 & 175 & 192 & $-78450$ & 0.29 & 3.08 & 0.998 \\
{BP4} &  & 486 & 486 & 543 & 31685  & 148 & 153 & 169 & $-56761$ & 1.98 & 7.38 & 0.999 \\
\hline
\end{tabular}%
}
\caption{Input parameters for the selected BPs in the I(1+2)HDM. All masses(squared) are in GeV$^{(2)}$.}
\label{BPs_descriptions}
\end{table}

The diagrams in Fig.~\ref{fig:generic_FD_I3HDM} show that the presence of additional (both active and inert) scalar states of the I(1+2)HDM with respect 
to the SM will manifest themselves differently depending on whether they are neutral or charged. Of the neutral ones, only the active $H$ state enters and this will  appear as an $s$-channel Breit-Wigner (BW) resonance, so long that $\sqrt s$ samples an interval containing $m_H$. Of the charged ones, both the active $H^\pm$ and inert $\chi^\pm$ states will enter  and these will appear as loop thresholds, so   long that $\sqrt s$ samples an interval containing $2m_{H^\pm}$ and $2m_{\chi^\pm}$, respectively.
%
  

\begin{figure}[!h]
\centering
\includegraphics[scale=0.4]{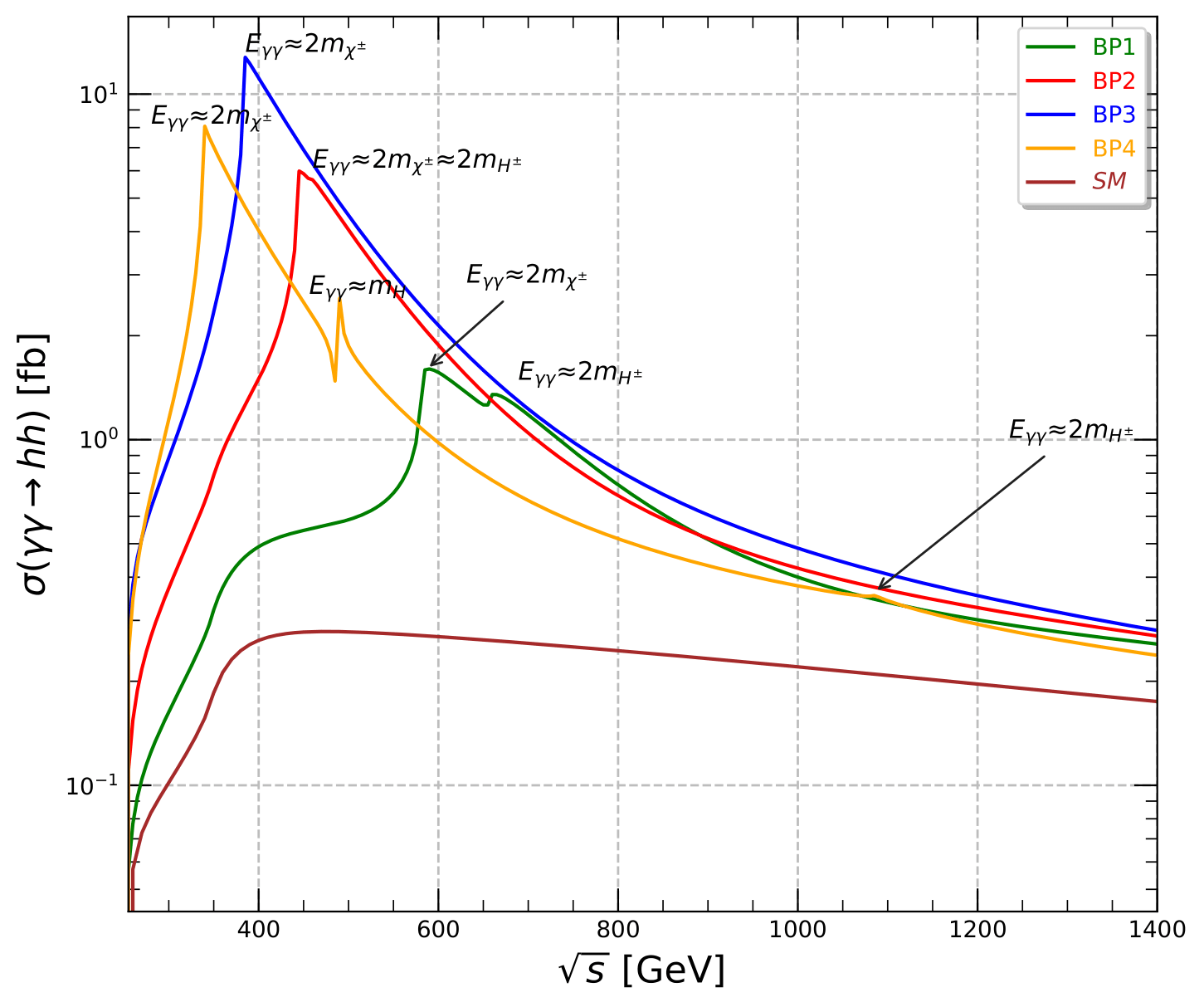}
\includegraphics[scale=0.4]{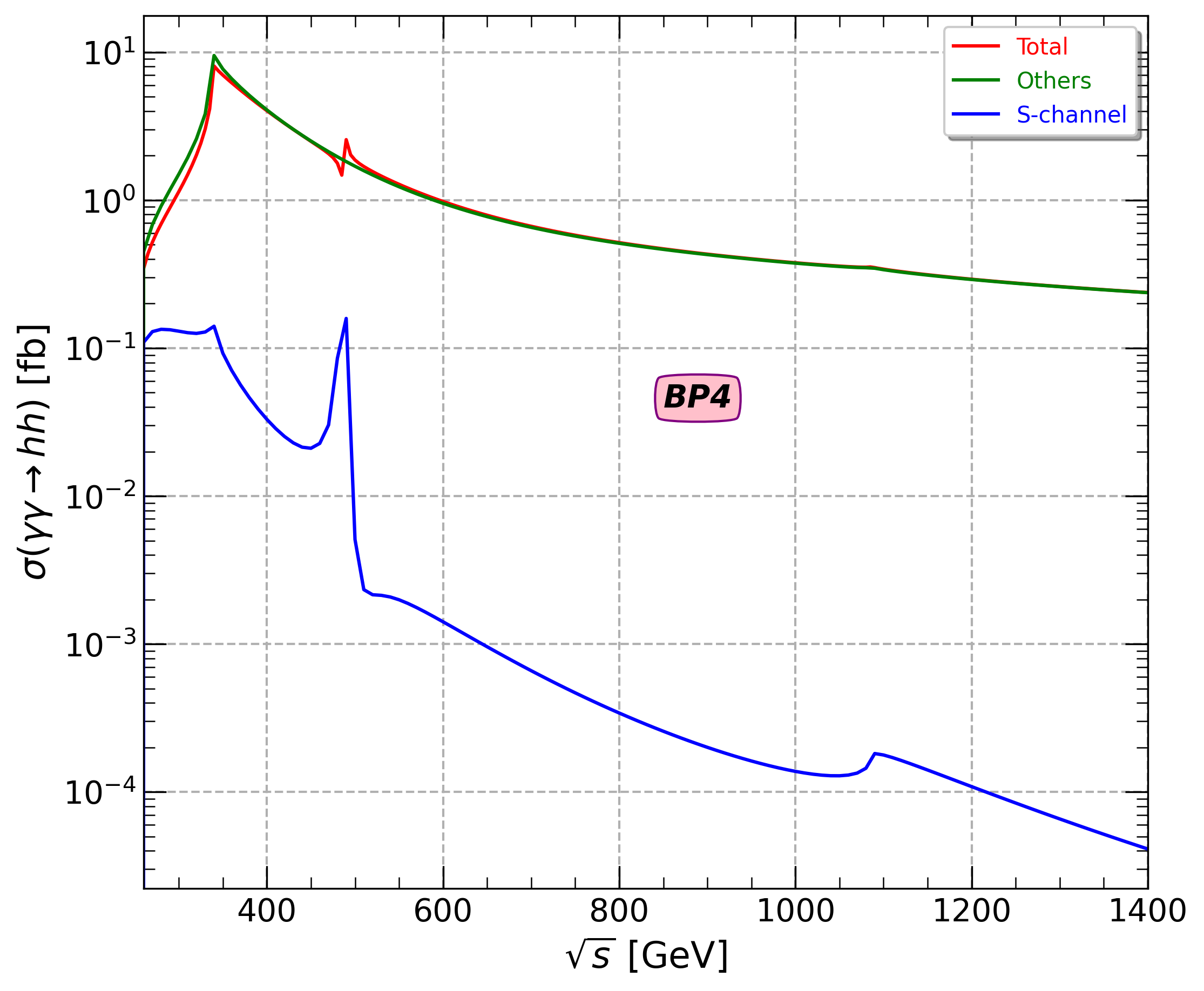}
\caption{Left: SM and I(1+2)HDM total cross section $\sigma(\gamma\gamma \to hh)$ as a function of the collision energy $\sqrt{s}$ for unpolarized beams for each BP. Right: I(1+2)HDM  cross section $\sigma(\gamma\gamma \to hh)$ as a function of the collision energy $\sqrt{s}$ for unpolarized beams for  BP4 decomposed in the two subchannels described in the text.}
\label{fig:XS_BPs}
\end{figure}

 All such roles are evident in Fig.~\ref{fig:XS_BPs} (left), where the total unpolarized partonic cross section $\sigma(\gamma\gamma \to  h h)$ is presented for 
 the selected BPs as a function of the CoM energy $\sqrt{s}$. Herein, 
the aforementioned threshold effect is observed for each BPs when $E_{\gamma\gamma} \approx 2 m_{\chi^{\pm}} $ or $2 m_{H^{\pm}}$, corresponding to the opening of one of the charged Higgs pair channels: $\gamma \gamma \rightarrow \chi^+ \chi^-$ or $ \gamma \gamma \rightarrow H^+ H^-$, respectively. All these kinematic configurations amplify the cross section, which reaches a maximum of about $12.77$ $fb$ for BP3 at $\sqrt{s} \approx 385$ GeV.   Similarly, a significant enhancement may occur in the inverted hierarchy case, where $m_{\chi^{\pm}} < m_{H^{\pm}}$ (BP4), leading to a cross section of approximately 8 $fb$ at lower $\sqrt{s}$. For BP1 and BP2, the enhancements with respect to the SM case are also evident. 
The BW resonance effect  is here only present for BP4, as only in this case $\sqrt s$ reaches $m_H$, {{so that  the heavy CP even (active) state $H$ can decay into $hh$. However, such a resonance is not expected to be very large given the fact that the $Hhh$ coupling is proportional to $\cos(\beta-\alpha)$,  which is very small as driven by LHC data (see Tab.~\ref{trilinear_quartic_couplings}).}}
More generally, the cross section for $\gamma\gamma\to hh$  in the I(1+2)HDM can attain substantially larger values compared to the SM prediction, by factors ranging from 
$10$ to almost $50$, hence, above and beyond what obtained by changing $\kappa_\lambda$ alone. We also note that, as the charged scalar masses increase, their loop effects decouple, hence causing the cross section to eventually drop, so as  to asymptotically approach the SM result. This behavior highlights the complementarity between direct searches for charged Higgs bosons and indirect probes via the $\gamma\gamma \to hh$ process.

\begin{table}[!ht]
\centering
\renewcommand{\arraystretch}{1.4}
\setlength{\tabcolsep}{8pt} 
\caption{Triple and quartic scalar  couplings in the I(1+2)HDM, normalized to their SM values, for our BPs.}
\begin{tabular}{lcccc}
\hline
Coupling ratio & {BP1} & {BP2} & {BP3} & {BP4} \\ 
\hline
$({Hhh})^{\model}/(hhh)^{\rm SM}$                & 0.014 & -0.002 & -0.056 & 0.42 \\
$({hhh})^{\model}/(hhh)^{\rm SM}$               & 0.97 & 0.97 & 0.97 & 0.95 \\
$({hH^+H^-})^{\model}/(hhh)^{\rm SM}$           & 4.07 & 1.86 & -0.36 & 3.88 \\
$({HH^+H^-})^{\model}/(hhh)^{\rm SM}$           & 1.44 & 1.49 & 0.41 & 0.04 \\
$({h \chi^+ \chi^-})^{\model}/(hhh)^{\rm SM}$   & 5.45 & 4.91 & 4.81 & 3.57 \\
$({H \chi^+ \chi^-})^{\model}/(hhh)^{\rm SM}$    & 0.60 & 0.37 & -0.27 & 0.12 \\ \hline
$({H^+ H^-hh})^{\model}/(hhhh)^{\rm SM}$        & 4.04 & 1.81 & -0.34 & 3.96 \\
$({\chi^+ \chi^- hh})^{\model}/(hhhh)^{\rm SM}$ & 5.59 & 5.01 & 4.91 & 3.64 \\
\hline
\end{tabular} 
\label{THC} 
\end{table}

Similarly to the SM, in the  I(1+2)HDM there is a destructive interference between the $s$-channel graphs and the others. This is illustrated in Fig.~\ref{fig:XS_BPs} (right)  where we can see that at low energy there is a slight reduction of the other contributions, while at high energy they fully dominate over the $s$-channel ones despite the appearance of  a resonant effect from $H\to hh$ decays.
Such an interference again leads to the characteristic pick-dip structure that appears just above the $hh$ kinematic threshold. As $\sqrt{s}$ increases, the other contributions decrease smoothly whereas the triangle terms rapidly fall off due to the $s$-channel propagator suppression. These results can be traced back to the structure of the scalar interactions in our NP model, in particular, the trilinear couplings $h\chi^+\chi^-$  and  $hH^+H^-$ together with the quartic couplings $hh\chi^+\chi^-$  and $hhH^+H^-$, which are relatively large, as can be seen from Tab.~\ref{THC}. These enhanced couplings amplify the other contributions compared to the $s$-channel ones, thereby shaping the overall energy dependence of the total cross section.

\begin{figure}[!ht]
\centering
\includegraphics[scale=0.4]{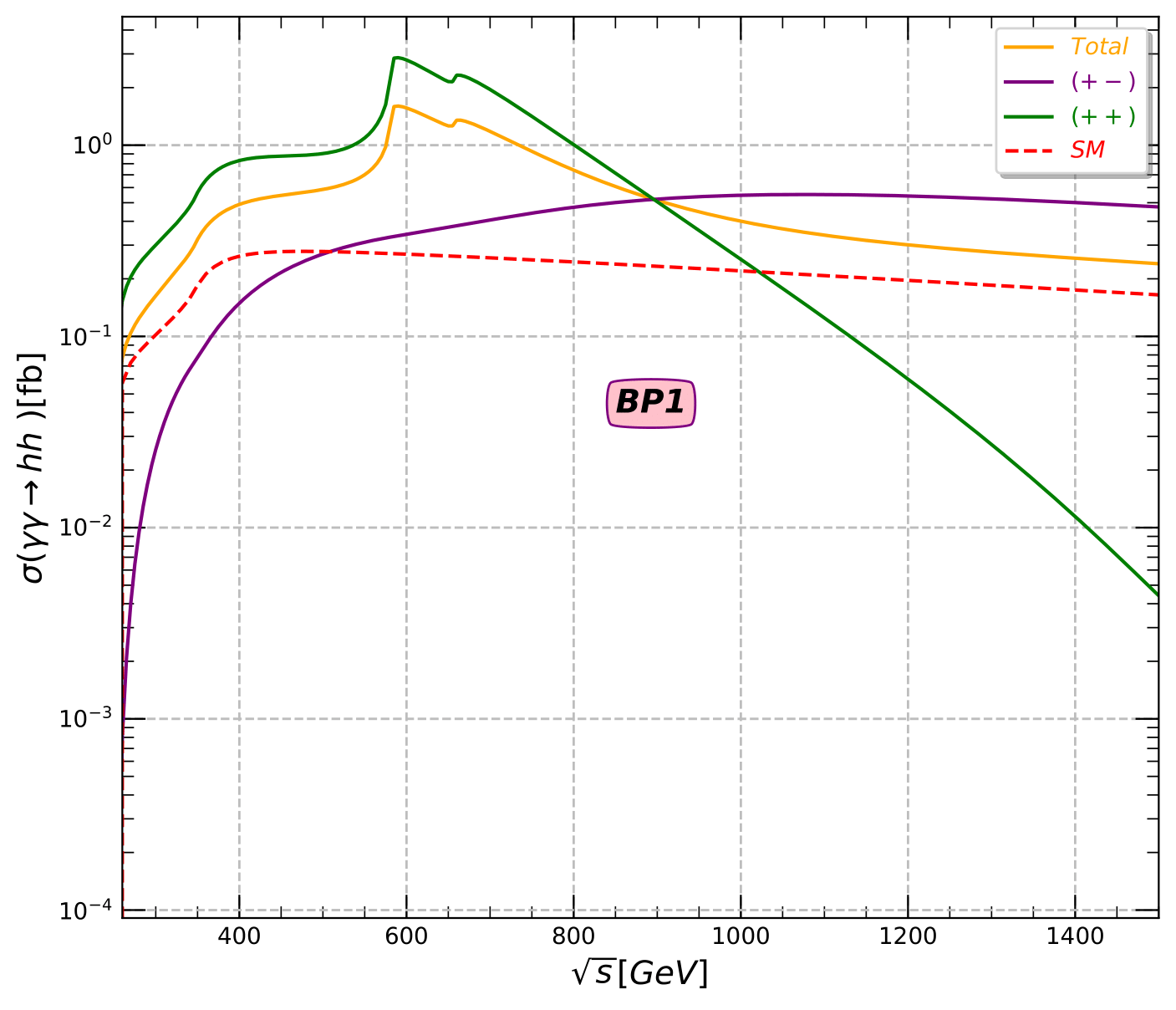}
\includegraphics[scale=0.4]{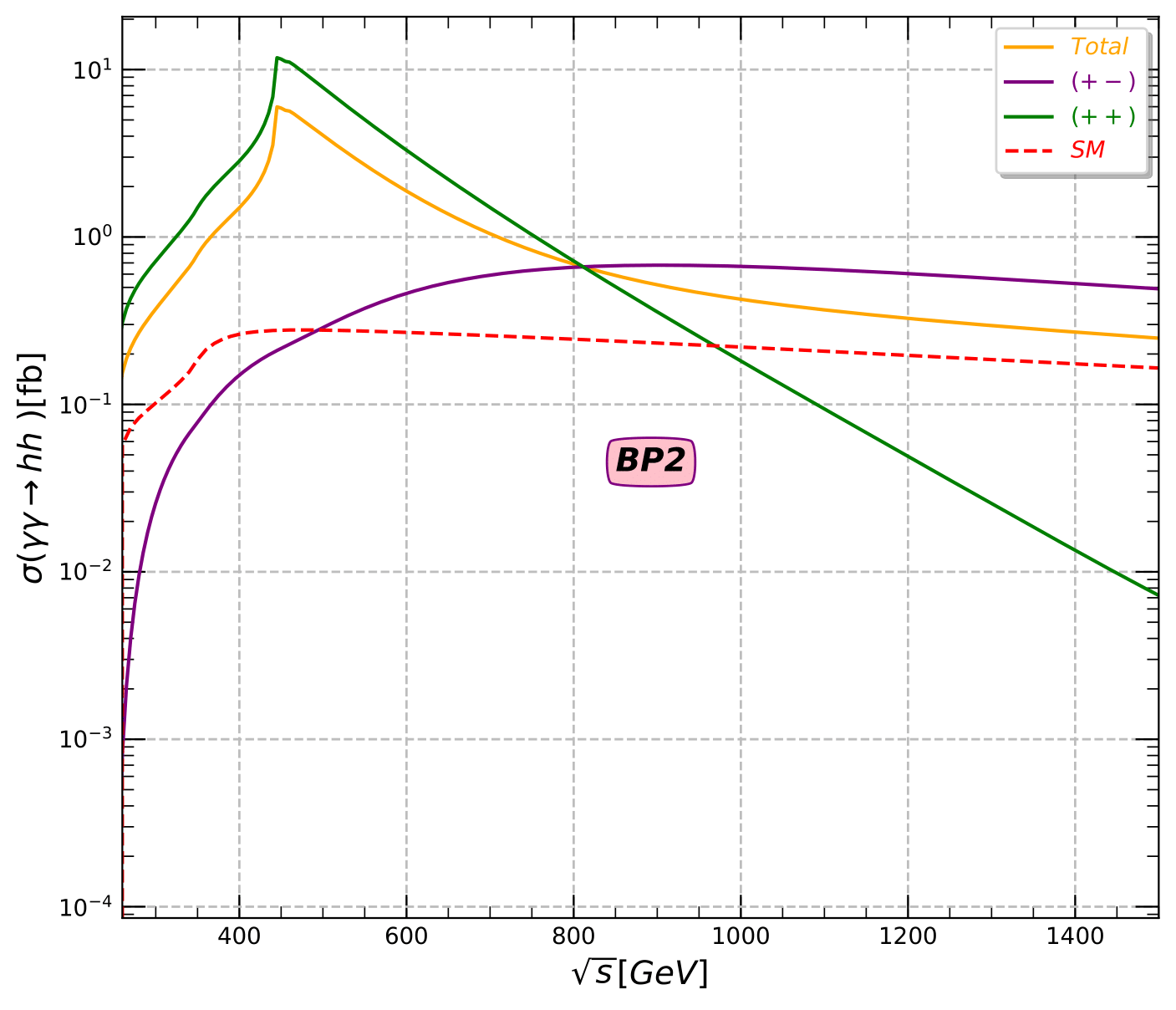}\\
\caption{SM and I(1+2)HDM total cross section $\sigma(\gamma\gamma \to hh)$ as a function of the CoM energy $\sqrt{s}$. For the I(1+2)HDM case, 
we also show the corresponding values for photons with opposite helicity $(+-)$ and identical helicity $(++)$ for BP1 (left) and BP2 (right).}
\label{fig:polarization}
\end{figure}

\begin{figure}[!h]
\centering
\includegraphics[scale=0.4]{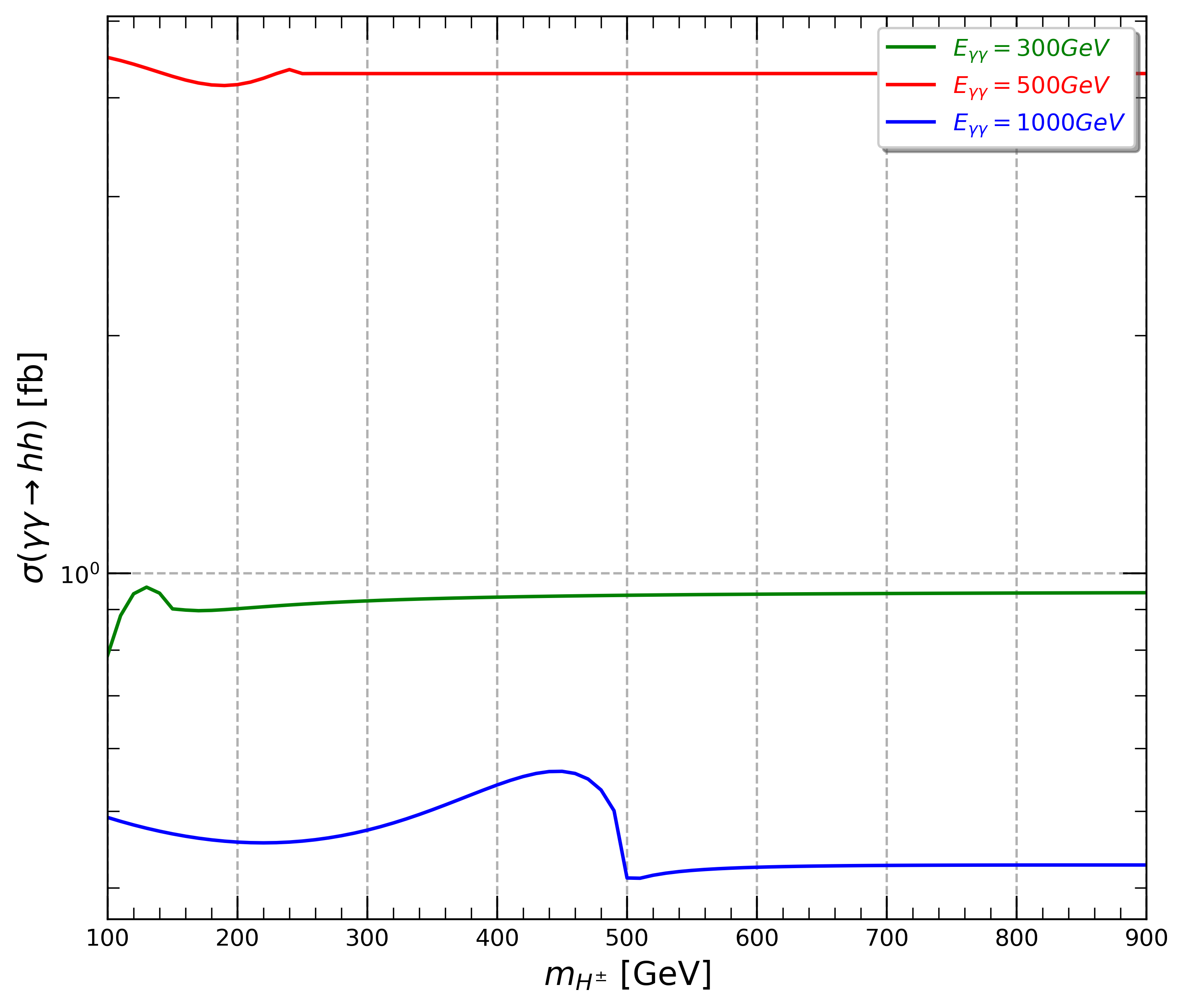}
\includegraphics[scale=0.4]{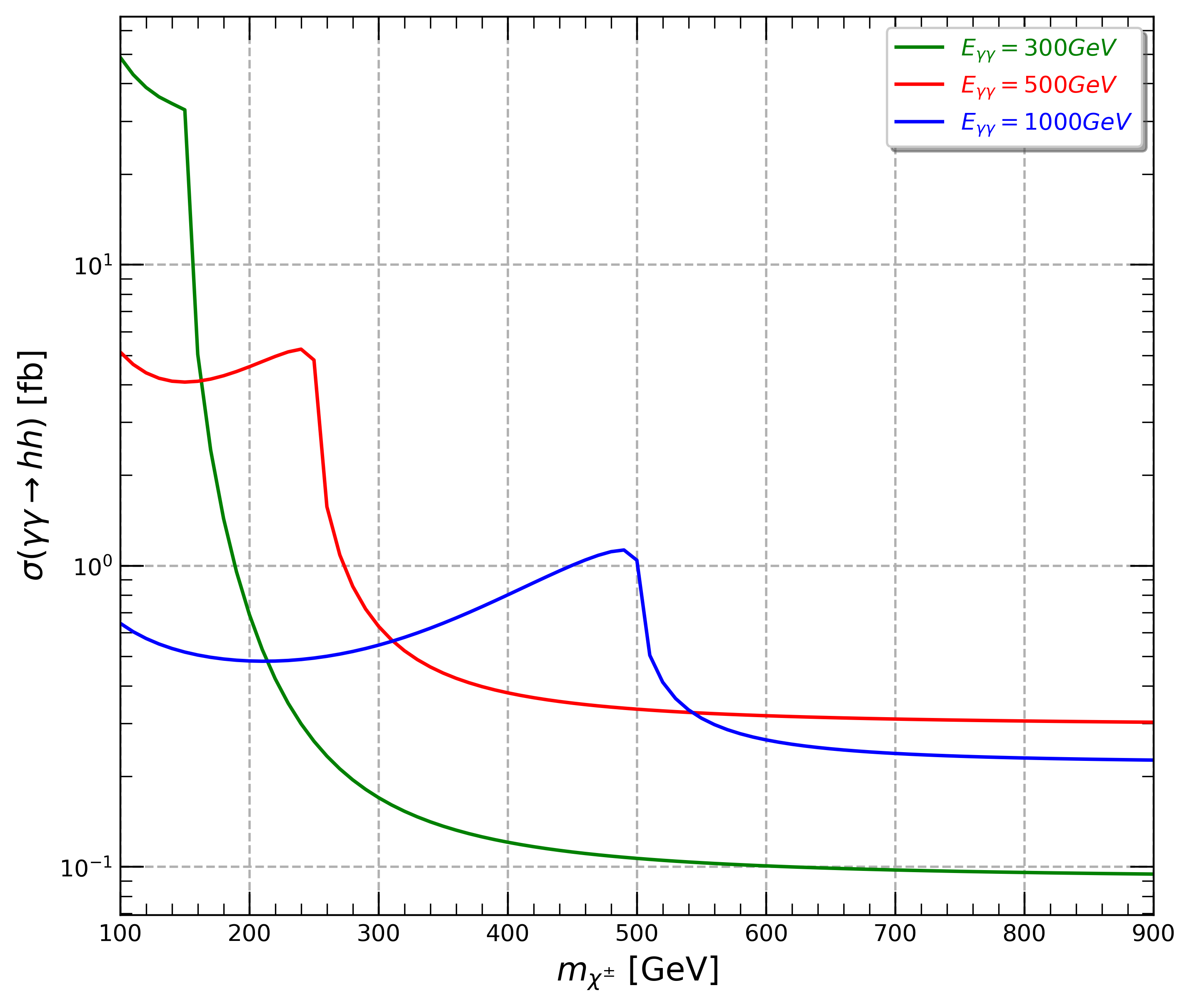}
\caption{I(1+2)HDM total cross section $\sigma(\gamma\gamma \to hh)$ as a function of the charged scalar masses, $m_{H^\pm}$ (left) and $m_{\chi^\pm}$ (right), for different values of $\sqrt{s}$. }
\label{fig:effect-ChipmHpm}
\end{figure}

Another, equally important aspect, in the study of the $\gamma\gamma\to hh$ cross section  involves the polarization of the incoming photon beams. In fact, exploiting the polarization of the photons is a vital technique for enhancing the sensitivity to the di-Higgs signal by suppressing backgrounds that stem from different helicity states of the photons. To address this issue in our study, we illustrate in Fig.~\ref{fig:polarization}, e.g., for BP1 and BP2,  the total cross section for the process $\gamma\gamma \to hh$ in both the SM and I(1+2)HDM as a function of the CoM energy $\sqrt{s}$, for different photon-photon helicity configurations\footnote{We define here the total angular momentum along the beam axis, $J_z$,  in terms of the helicities of the two incoming photons ($i=1,2$) as follows: $J_z=|\lambda_{\gamma}^1-\lambda_{\gamma}^2|$, where $\lambda_{\gamma}^i=\pm1$.}. The orange curve represents the unpolarized case while the purple and green lines correspond to the opposite helicity $(+-)$ and same helicity $(++)$ photon states, respectively. Interestingly, the $J_z=0$ configuration $(++)$ provides the dominant contribution to the di-Higgs production rate, since it couples to the CP-even component of the effective $\gamma\gamma hh$ interaction generated by loop diagrams involving charged particles. In contrast, the $J_z=2$ configuration $(+-)$ is  suppressed and becomes more relevant only at higher energies due to non-$s$-channel loop contributions. In this context, and for all the BPs considered, the $(++)$ channel exhibits in the I(1+2)HDM a noticeable enhancement relative to the SM, especially near the various threshold and resonance regions, reflecting the impact of modified self-interactions and additional charged scalars. 

Concerning this last point, Fig.~\ref{fig:effect-ChipmHpm} illustrates the sensitivity of the total cross section to the charged scalar masses, $m_{H^\pm}$ (left) and $m_{\chi^\pm}$ (right), for several values of the CoM energy $\sqrt{s}$. As can be seen, the corresponding cross section exhibits a pronounced enhancement, particularly near the kinematic thresholds, followed by a rapid suppression once these  are crossed, i.e., $m_{H^\pm},\,m_{\chi^\pm} \gtrsim \sqrt{s}/2$. This behavior reflects the opening and subsequent decoupling of the charged scalar pair production channels in the loops, $\gamma\gamma \to H^+H^-,\,\chi^+\chi^-$, which strongly influence the loop-induced di-Higgs production rate.

\begin{figure}[!h]
\centering
\includegraphics[scale=0.4]{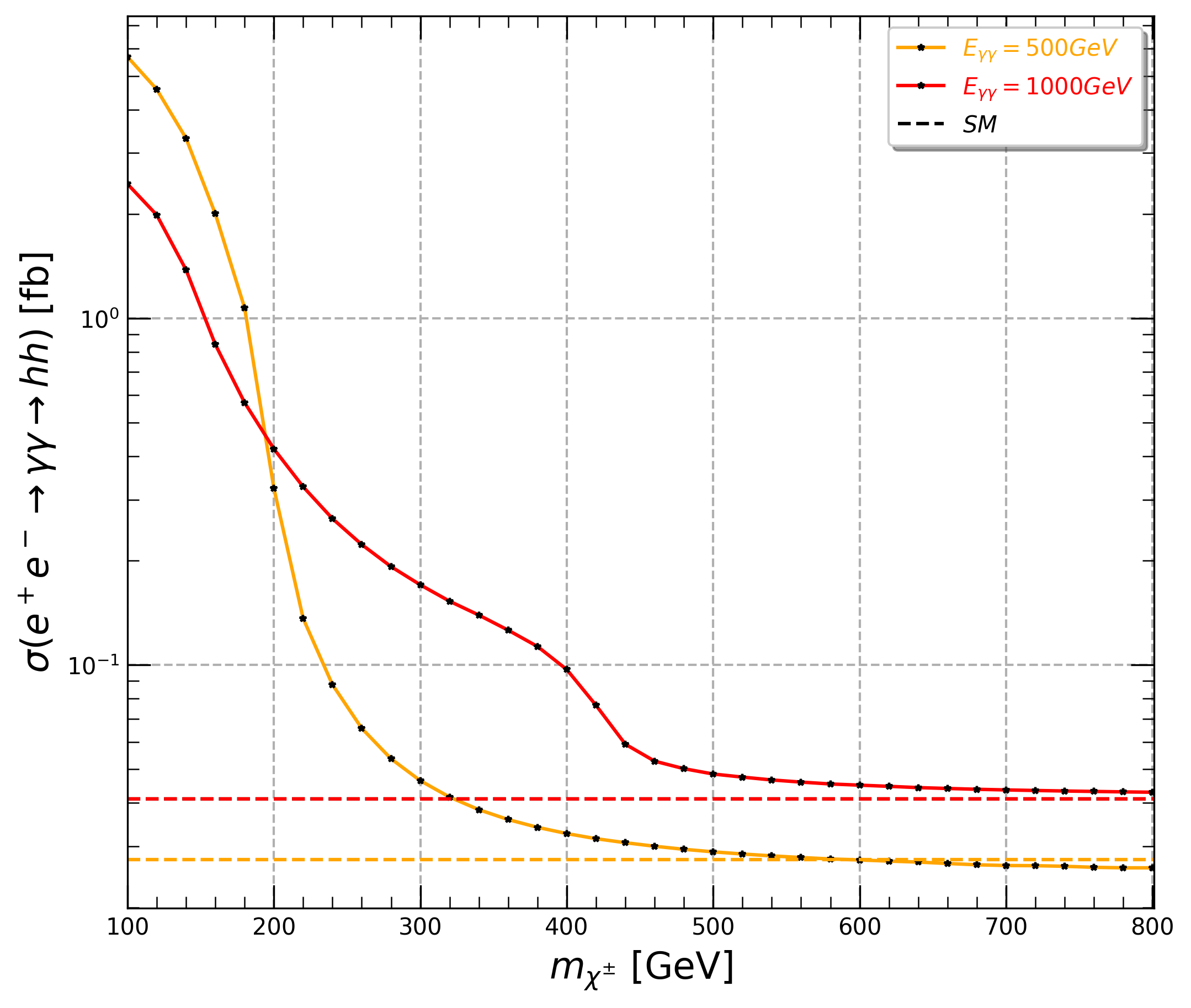}
\includegraphics[scale=0.4]{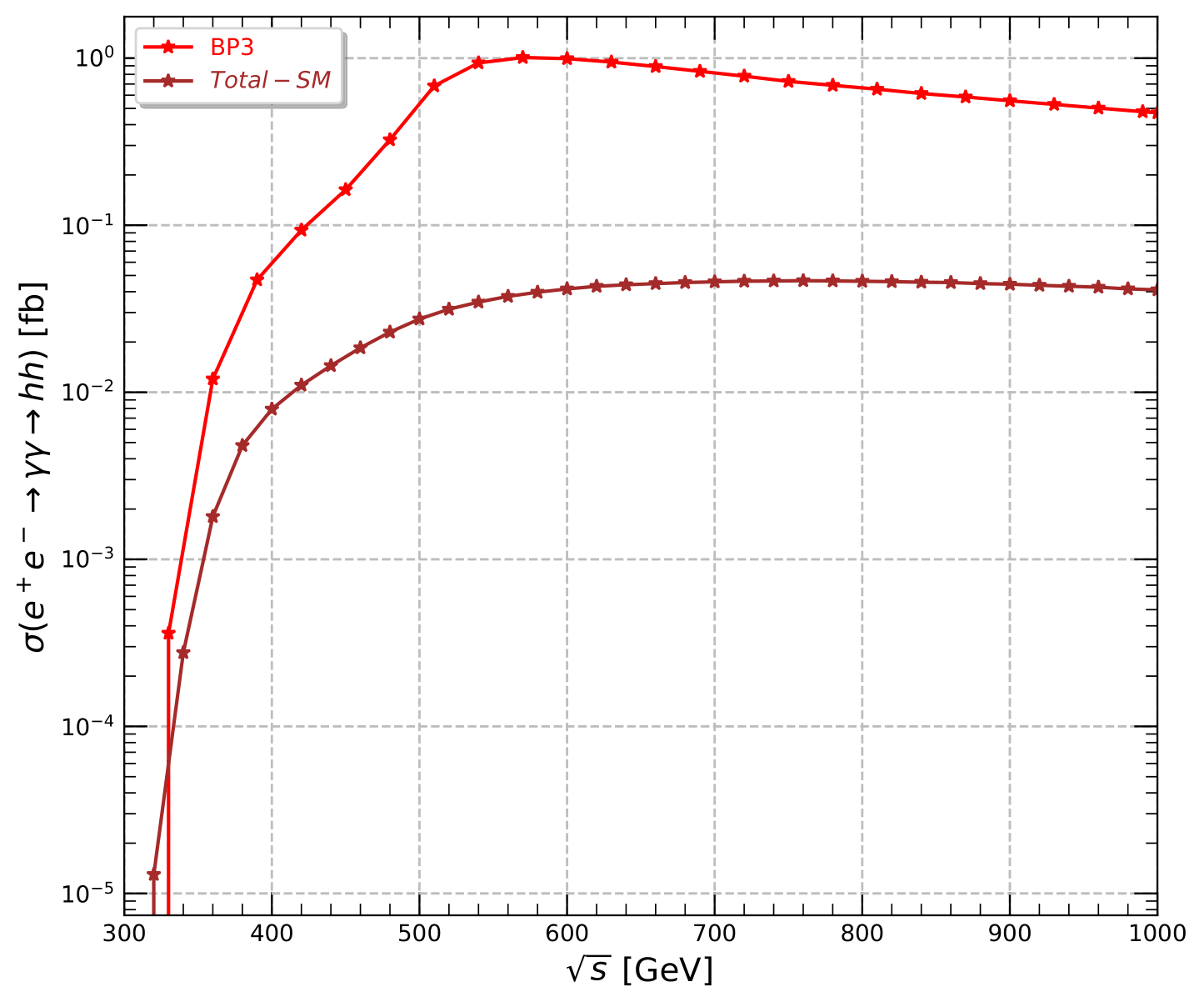}
\caption{The total  cross section $\sigma(e^+e^-\to\gamma\gamma \to hh)$ as a function of the  inert charged scalar mass $m_{\chi^{\pm}}$ (left) and as a function of the collision energy $\sqrt{s}$ (right) for unpolarized beams in the case of BP3.}
\label{fig:xs-convoluted}
\end{figure}

So far, we have treated the two  photons in the initial state as real objects. In reality, being produced via Compton back-scattering of laser light, their energy distribution is different from that of the actual incoming beams of electrons and positrons\footnote{Note that Quantum Electro-Dynamics (QED) interactions preserve helicity, so that the polarization of the $e^\pm$ beams is transferred to the photons.}.
Thereafter, we account for this effect by convoluting our previous partonic results with the photon-photon luminosity function (as previously described).
Specifically, we show in the left panel of Fig.~\ref{fig:xs-convoluted} how the cross section for $e^+ e^- \to \gamma\gamma \to hh$ decreases with increasing $m_{\chi^\pm}$ for whatever $\sqrt{s}$, e.g., in the case of BP3. 
(The SM cross section as a function of $\sqrt{s}$ can be read from the right panel of the same figure.) For instance, at $\sqrt{s}=500\,\text{GeV}$ and $m_{\chi^{\pm}}=100\,\text{GeV}$, we see that the I(1+2)HDM cross section reaches a value of 5.68 $fb$. Thus, the magnitude of this rate is comparable to, or even significantly exceeds, the SM expectation in certain areas of parameter space, notably, for higher values of $\sqrt{s}$ (see right panel of Fig.~\ref{fig:xs-convoluted}).

\begin{figure}[!h]
\centering
\includegraphics[scale=0.25]{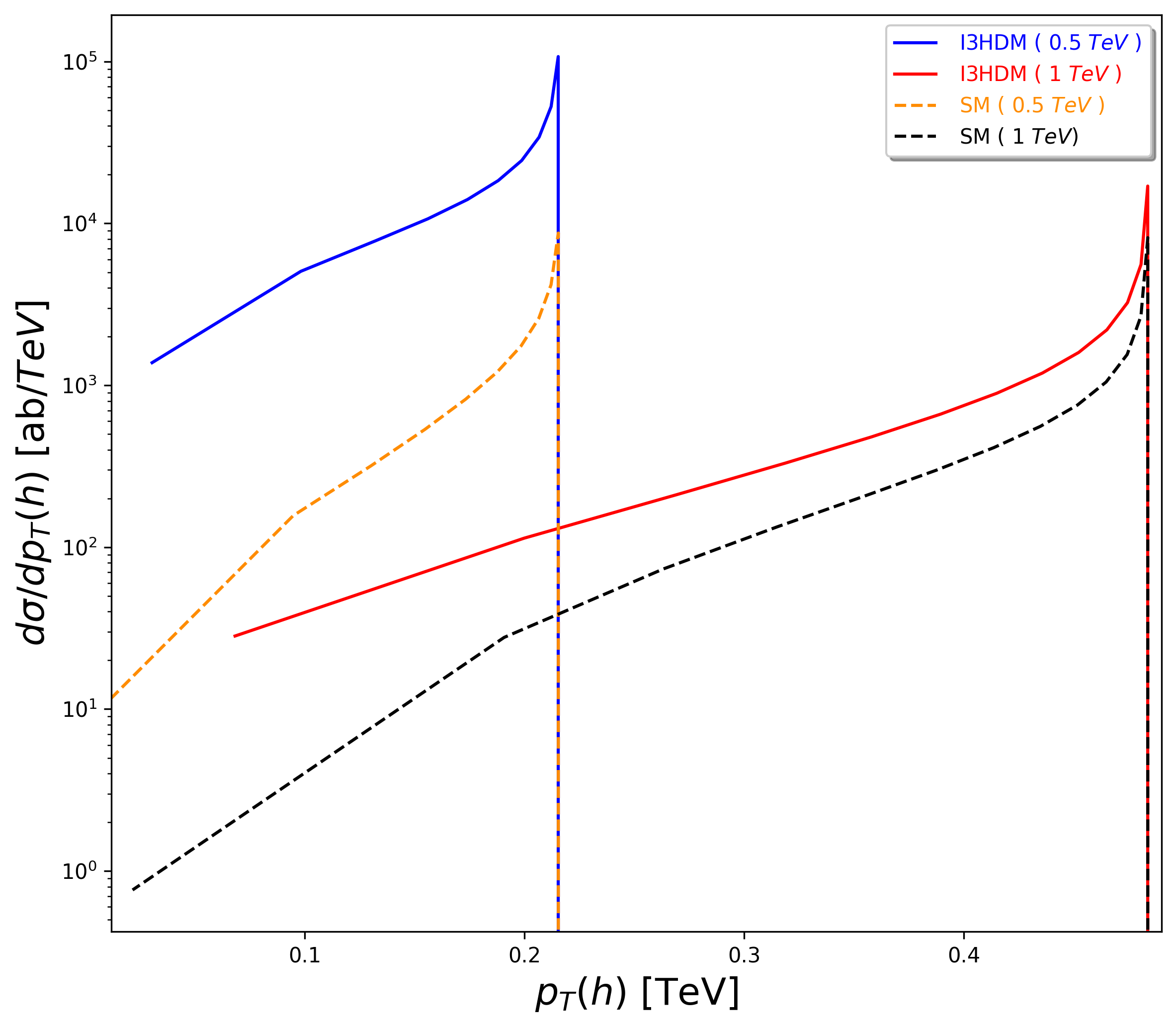}
\includegraphics[scale=0.25]{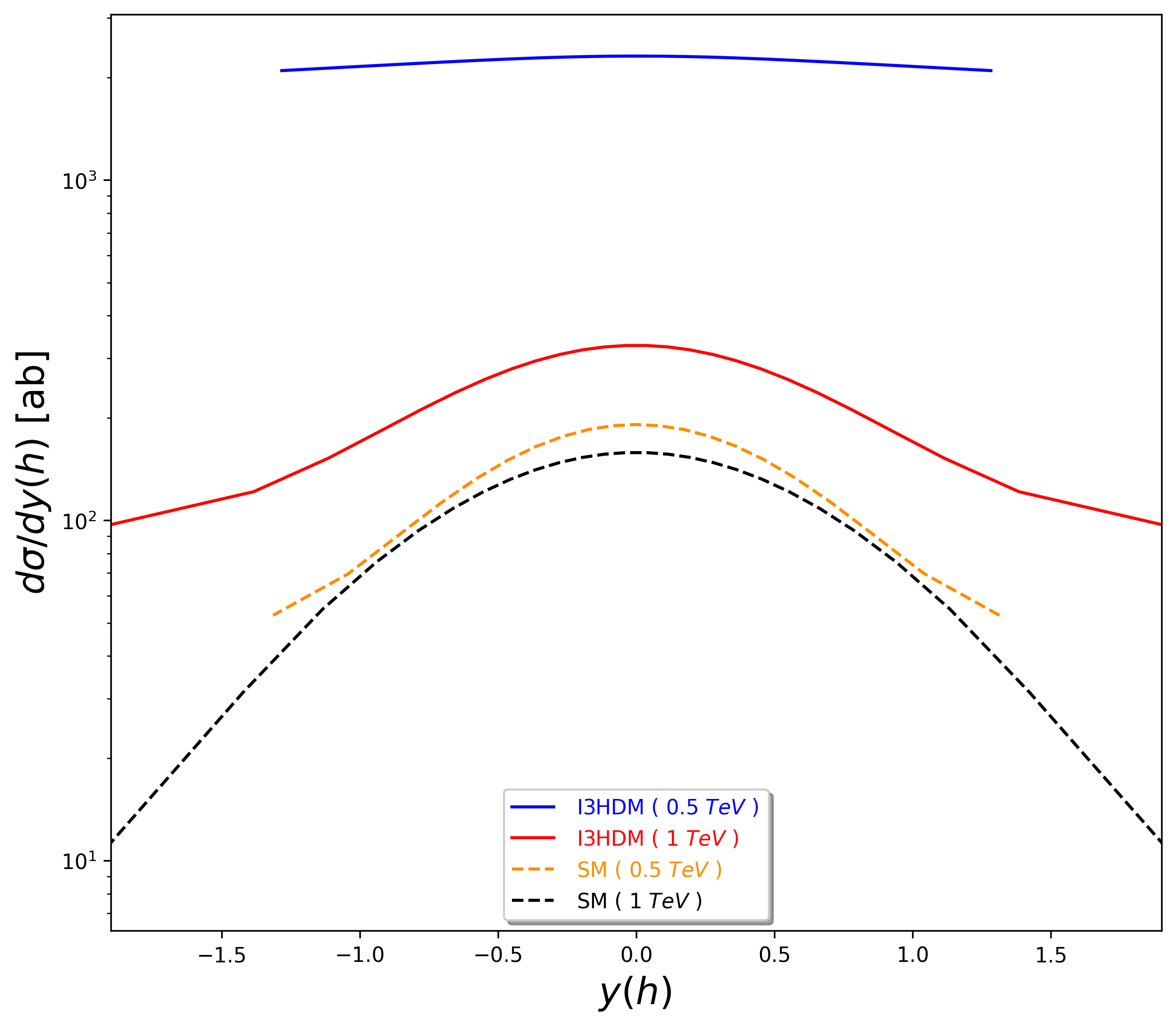}
\includegraphics[scale=0.25]{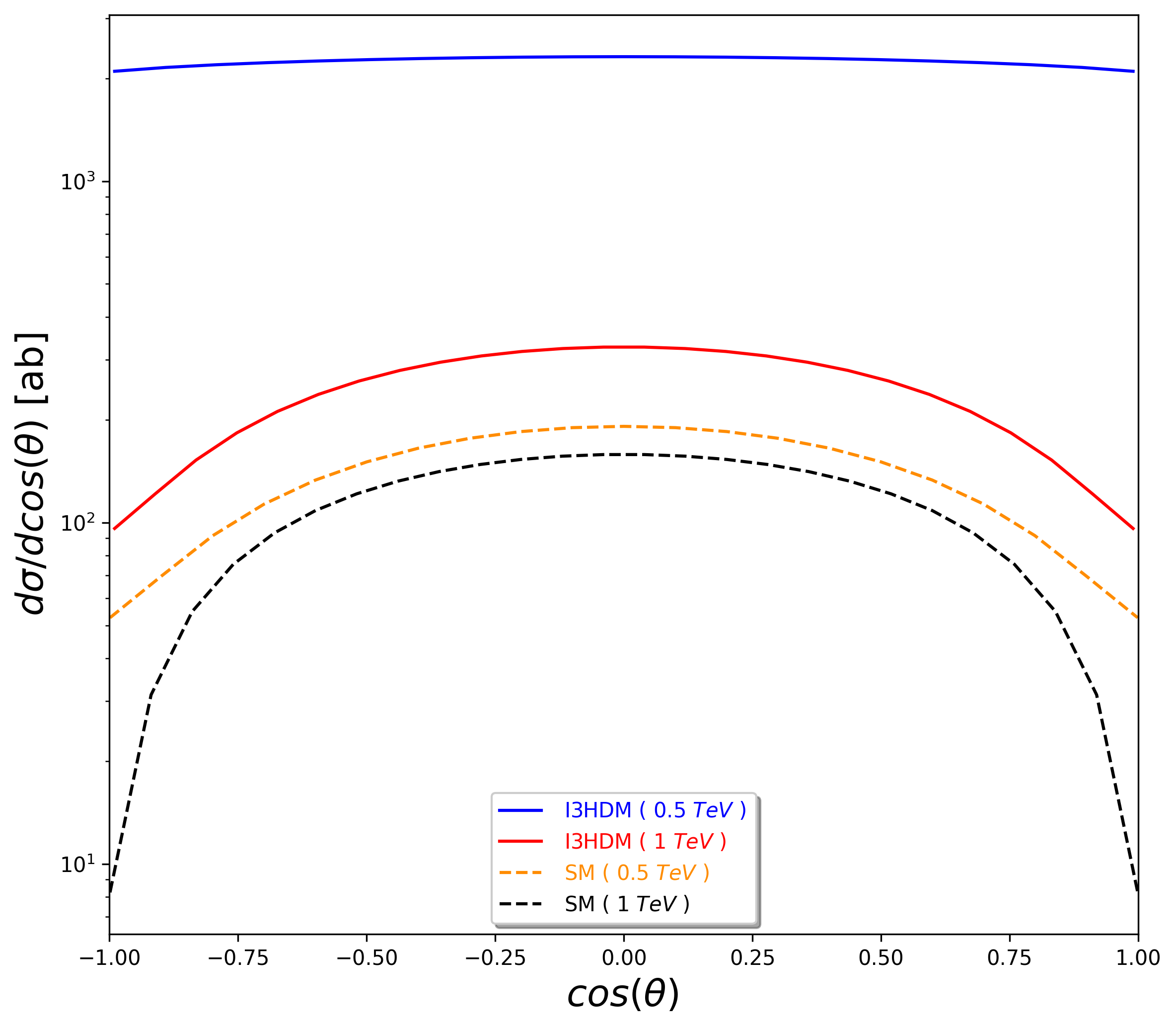}
\caption{The transverse-momentum (left), rapidity (center) and polar angle  (right) distributions of either $h$ state in the process $\gamma\gamma \to hh$ within the I(1+2)HDM, for two values of the $\sqrt{s}$, 500 GeV and 1000 GeV, for BP3.}
\label{fig:distributions}
\end{figure}
To end our study, we discuss some kinematical distributions for the partonic di-Higgs production, in the $\gamma\gamma$ CoM system, before the convolution with the photon energy spectrum emerging from the $e^+e^-$ beams. (The results for these observables are not drastically different between the two descriptions.)
The three panels of Fig.~\ref{fig:distributions} display the differential cross section as a function of the SM-like Higgs transverse-momentum $p_T(h)$ (left), rapidity $y(h)$ (middle) and polar angle $\cos\theta$ (right), for two representative CoM energies, $\sqrt{s}=500~\text{GeV}$ and $1000~\text{GeV}$.  We note here that both Higgs bosons have the same energy $E_{h}=\sqrt{s}/2$ and are in a back-to-back configuration (as we are not convoluting the photon beams with the electron/positron structure functions). 
As a result, by comparing the SM predictions with the I(1+2)HDM findings, here illustrated for BP3, 
it is clearly visible that the transverse momentum distribution exhibits a characteristic rise near the production threshold, followed by a sharp decrease at higher $p_T(h)$ values, as expected from the kinematic limit at $\sqrt s/2$. The I(1+2)HDM curves consistently lie above the SM ones, particularly at lower $p_T(h)$, where the enhancement arises from additional charged scalar loop effects. The rapidity distribution  peaks in the central region, indicating that most SM-like Higgs bosons are produced in the transverse direction with respect to the beam axis. A similar feature is obtained in the angular distribution, which is forward-backward symmetric, as required by CP conservation in the scalar sector, which remains nearly flat, thus indicating that the $\chi^{\pm}$ and  $H^{\pm}$ scalar contributions dominate over the SM and substantially reduce the angular dependence. This effect is most pronounced at $\sqrt{s}=500\,\text{GeV}$, but still  visible at 1000 GeV, thus highlighting also at  the differential level the sensitivity of the $\gamma\gamma \to hh$ process to both charged scalar sectors.

\section{Summary}\label{sec:summary}

We have carried out studies of pair production of SM-like Higgs bosons at future $\gamma\gamma$ linear colliders, wherein photons are obtained via  Compton back-scattering of laser light from  $e^+e^-$ beams,  within the framework of the I(1+2)HDM, a distinctive BSM scenario that includes both active (Higgs) and inert charged scalar (i.e., spin-0) states entering such a process at lowest order, alongside the SM particles. Crucially, 
since the $\gamma\gamma\to hh$ process is mediated at one-loop level in the SM, this is an ideal place to look for NP effects, as the latter enter at the same perturbative order as the former. The relevant one-loop amplitudes have been computed here in the Feynman gauge using dimensional regularization and the intervening graphs in both the SM and I(1+2)HDM have been grouped in such a way to render the underpinning physics most evident. 

In evaluating both the SM and I(1+2)HDM contributions, we have emphasized, in particular,  the effects of  the extra charged particles of   our BSM scenario. This has been done in part because both of the latter, the customary $H^\pm$ states of the active 2HDM sector and the new $\chi^\pm$ states of the inert sector,  play a significant role in explaining current data anomalies in Higgs boson searches. However, above and beyond this present motivation, we were keen to assess whether it would be possible to see their presence in the cross section of the above process as loop thresholds, given the ability that $\gamma\gamma$ colliders have of  scanning over the beam energy with high precision. With this in mind,  
 numerical results have been presented for  the (partonic) cross section $\gamma\gamma \to hh$ as well as the total (photonic) one $e^+e^- \to \gamma \gamma \to hh$.

After imposing theoretical and experimental constraints, we have found in the I(1+2)HDM enhancements of up to 2 orders of magnitude compared to the SM, with, indeeed, very characteristic effects near the charged scalar pair production thresholds which offset the underlying  destructive interferences between the $s$-channels and other diagrams. 
{{Furthermore, we have also shown that the amplitude $\gamma \gamma \to hh$  would have a $(hS^+S^-)^2$ ($S^\pm=\chi^\pm,H^\pm$) dependence through the the latter.  This would in turn enable one to extract the value of $m_{H^\pm}$, $m_{\chi^\pm}$ and also $hS^+S^-$,}} so that, if  the HL-LHC can extract the triple Higgs coupling modifier $\kappa_\lambda$ entering the $hhh$ interaction, one could use  $\gamma \gamma \to hh$ production 
to shade some information on the trilinear scalar self-couplings $hH^+H^-$ and $h\chi^+\chi^-$ better than one could do in $h\to\gamma\gamma$ decay while also been able to access the  $hhH^+H^-$ and   $hh\chi^+\chi^-$ couplings, which are precluded at the decay level. Finally, we have shown that the polarization of the initial electron/positron (and, in turn) photon beams further increases the I(1+2)HDM cross sections relative to the SM. 

All these features thus make $\gamma\gamma \to hh$ an excellent channel to test both trilinear and quartic scalar  
self-couplings and to determine the charged scalar spectrum with high precision. Specifically,  
our results highlight the strong potential of future $\gamma\gamma$ colliders to uncover distinctive signatures of extended Higgs sectors, such as in the I(1+2)HDM, at future electron-positron linear colliders: in particular,   we have presented numerical results here applicable to the ILC and CLIC prototypes
at both the integrated and differential level.

\section*{Acknowledgements}
AA is supported by the Arab Fund for economic and social development.  SM is supported in part through the NExT Institute and  STFC Consolidated Grant No.  ST/X000583/1. 
LR would like to thank the CERN Department of Theoretical Physics for its hospitality and stimulating environment, where a part of this work was carried out.

\bibliographystyle{JHEP}
\bibliography{draft}

\end{document}